\documentstyle[12pt]{article}
\begin{document}
                 \def\gappeq{\mathrel{\rlap {\raise.5ex\hbox{$>$}} 
                 {\lower.5ex\hbox{$\sim$}}}}
                 \def\lappeq{\mathrel{\rlap{\raise.5ex\hbox{$<$}}
                  {\lower.5ex\hbox{$\sim$}}}}
         \def \gsim{\lower.8ex\hbox{$\sim$}\kern-.75em\raise.45ex\hbox{$>$}\;}
         \def \lsim{\lower.8ex\hbox{$\sim$}\kern-.8em\raise.45ex\hbox{$<$}\;}
                      \def\ltsima{$\; \buildrel < \over \sim \;$}
                      \def\simlt{\lower.5ex\hbox{\ltsima}}
                       \def\rtsima{$\; \buildrel > \over \sim \;$}
                      \def\simrt{\lower.5ex\hbox{\rtsima}}
          \def\gappr{\lower 3pt\hbox{$\buildrel > \over \sim\;$}}
          \def\gappl{\lower 3pt\hbox{$\buildrel < \over \sim\;$}}
\centerline{\bf Eikonal profile functions and amplitudes }
\centerline{\bf  for $\rm pp$ and $\bar{\rm p}{\rm p}$ scattering }
\centerline{Fl\'avio Pereira}
\centerline{Departamento de Astronomia Gal\'actica e Extragal\'actica}
\centerline {Observat\'orio Nacional, CNPq, Rio de Janeiro 20921-400, RJ, Brazil }
\centerline{ Erasmo Ferreira}
\centerline{ Instituto de F\'\i sica, Universidade Federal do Rio de Janeiro,}
\centerline{  Rio de Janeiro 21945-970, RJ, Brazil}

\begin{abstract}
 The eikonal profile function $J(b)$ obtained from the Model of the 
Stochastic Vacuum is parametrized in a form suitable for comparison 
with experiment. The amplitude and the extended profile function
(including imaginary and real parts) are determined directly from the
complete pp and $\bar {\rm p}$p elastic scattering data at high energies.
Full and accurate representation of the data is presented, with 
smooth energy dependence of all parameters. The changes needed in 
the original profile function required for description of scattering 
beyond the forward direction are described.
\end{abstract}
\vspace{2 cm}

\centerline{To appear in Physical Review D58}
  \newpage
 
\bigskip

{\bf 1. Introduction }

The application of the Model of the Stochastic Vacuum (MSV) 
\cite{RRG3,RRG3A,RRG3B} to pp and $\bar {\rm p}$p scattering 
gives a good description of the experimental quantities
related to the amplitude in the forward direction. 
For large values of the momentum transfer $|t|$ the experimental 
amplitude has a delicate structure, with a characteristic dip in the 
differential elastic cross section that is not reproduced by MSV
or  other dynamical model without free parameters.
 It is well known that the profile function of the eikonal
formalism, related to the
amplitude through a 2-dimensional Fourier transform, has an apparently 
structureless 
shape, and slight changes in the profile may cause dramatic
changes in the amplitude. The main reason is that the dominating 
imaginary part
of the amplitude passes through  zero for a value of $|t|$ which is not
large, while the real part, being comparatively small for low $|t|$, fills 
partially the dip due to the zero of the imaginary part, thus producing
the remarkable structure of the differential cross section.
 This is not really a shortcoming of MSV, since this model, in its
present formulation using the first term of an expansion based on the 
eikonal formalism, has not the purpose 
of describing the differential cross section much beyond the $|t|=0$ limit. 

Our study of the MSV profile function suggests analytical 
forms which can be used, with appropriate changes of parameters, 
to describe the data in the full range of $|t|$ values. 
In order to identify the necessary changes, we first describe 
accurately the whole data of high energy pp (at the ISR energies
$\sqrt{s}$= 20-60 GeV) and $\bar {\rm p}$p (at $\sqrt{s}$= 540 and 1800 
GeV) scattering in terms of an amplitude built with a  few terms
controlled by a number of parameters. This part of our work is
similar to the treatment of the ISR data by Carvalho and Menon 
\cite{MENON1}, but our parameters are fewer  and  have a much more 
regular energy dependence (which is essential to allow for predictions 
through interpolations and extrapolations). The comparison of the two  
different parametrizations exhibits
some freedom of  analytical forms allowed by the data. 

  The use of an impact picture representation for the scattering amplitude
of pp and  $\bar {\rm p} {rm p}$ scattering was first made by Bourrely,
Soffer and Wu \cite{WU}. With a $t$ dependence inspired in the
proton electromagnetic form factor, using a total of 6 adjustable 
parameters, they are able to reproduce the general features of the 
 ISR experiments. The form of the energy dependence is based on 
general properties of   quantum field theory, and has allowed 
good predictions por the CERN measurements at 540 GeV. The difference 
of  approach with respect to our work is that we take different forms
of parametrization (inspired on the  model of the stochastic vacuum)
 and that we do not parametrize the energy dependence. We show that 
very accurate fitting at every energy can be obtained with very smooth
energy dependence for the parameters. Our description includes the 
data at very high $|t|$ (up to 15 GeV$^2$) valid for all ISR energies
and we remark that the sign of the term describing this tail is 
reponsible for differences between pp and $\bar {\rm p}{\rm p}$ data
in this range. In the work of Bourrely, Soffer and Wu this large $|t|$ 
tail would be described by the Regge background term.
  
 We attempt to obtain  from the shapes of the eikonal functions some 
understanding of the geometric characteristics of 
the systems in collision, and thus try to visualize from the data the 
meaning of  effective radii for hadrons in collision\cite{RRG3D}.

\vspace*{5cm} 

 \centerline {  Table I - Experimental data}
\bigskip

  Table I shows the  available data on 
total cross sections, slope parameter and ratio of the forward real 
to imaginary parts of the amplitude in pp scattering, which 
come from  the  CERN Intersecting Storage Rings (ISR) 
measurements [\cite{RRDAT1}a,b,c] at energies ranging from $\sqrt{s}=23~\rm 
GeV$ to $\sqrt{s}=63~\rm GeV$, and in $\bar {\rm p}$p scattering at 
the energies $\sqrt{s}=541-546~\rm GeV$ 
 from the  CERN Super Proton Synchrotron (SPS) 
[\cite{RRDAT1}d] and the $\sqrt{s}=1800~\rm GeV$ 
information [\cite{RRDAT1}e] 
from Fermilab.  In addition, there are the data on differential 
cross sections in the range $\sqrt{s}$=23 - 63 GeV 
[\cite{RRDAT1}a],\cite{RRDAT2}, and  
for $\bar {\rm p}$p at $\sqrt{s}=546~\rm GeV$ \cite{RRDAT3} 
and $\sqrt{s}=1800~\rm GeV$ \cite{RRDAT6}.
We also give particular attention to the behavior of the amplitude for 
large $|t|$ measured  \cite{RRDAT5} at $\sqrt{s}=27~\rm GeV$. The data 
from Fermilab CDF \cite{ABE} at 1800 GeV giving $\sigma^T=$80 mb seem 
discrepant 
when we consider the smoothness of the energy dependence of our 
parametrization, and also is not in agreement with the predictions 
of Bourrely, Soffer and Wu \cite{WU}, and thus in the present work we 
concentrate on the results of the Fermilab E-710 experiment given in 
Table I.

  In the MSV calculation hadronic structures enter through transverse 
two-dimensional wave-functions in the plane perpendicular 
to the direction of the colliding hadrons, for which is made the simple 
ansatz
\begin{equation}
     \psi _{H} (R)=  \sqrt {2/\pi}\frac {1}{S_H} \exp{(-R^{2}/S_H^{2})}~,
\label{4A12}
\end{equation}  
 where $S_H$  is a parameter for the hadron size.
The dimensionless scattering amplitude $T_{H_1H_2}$ in the eikonal
formalism  is given in terms 
of the dimensionless profile function $\widehat J_{H_1H_2}$ for hadron-hadron 
scattering by 
\begin{equation}
 T_{H_1 H_2} = i s [\langle g^2 FF\rangle a^4]^2 a^{2}\int d^2 \vec b~
          \exp{(i \vec q\cdot \vec b )}~\widehat J_{H_1 H_2}
 (\vec b,S_1,S_2) ~,
 \label{4A14}
 \end{equation}
  where the impact parameter vector  $\vec b$  and the hadron sizes
  $S_1$, $S_2$  appear only  in units of the
  correlation length $a$, and $ \vec q $ is the momentum
  transfer projected on the transverse plane, in units of $1/a$,
  so that the momentum transfer squared is $t=-|\vec q|^2/a^2$. 
   The quantity $ \langle g^2 FF\rangle $  represents the gluon condensate.
  For short, from now on we write $J(b)$ or $J(b/a)$  to represent
 $\widehat J_{H_1H_2}(\vec b,S_1,S_2)$.
        The normalization of $T_{H_1 H_2}$ is such that total
and differential cross sections are given by 
\begin{eqnarray}
 \sigma^T  = \frac{1}{s}~\hbox{\rm Im}~T_{H_1 H_2} ~~~,~~~
    \frac {d\sigma^{e\ell}}{dt} =
             \frac {1}{16\pi s^2}~\vert T_{H_1 H_2}\vert^2  ~.
\label{4A15}
\end{eqnarray}
        To write convenient expressions for the observables, we 
 define the dimensionless moments of the profile function 
(as before, with $b$ in units of the correlation length $a$)
\begin{equation}
              I_k = \int d^2\vec b~b^k~ J(b) ~~,~k=0, 1, 2, ...
\label{4A18}
\end{equation}
which depend only on $ S_1/a$, $ S_2/a$, and the Fourier-Bessel
 transform
\begin{equation}
              I(t) = \int d^2\vec b~ J_0 (b a \sqrt{|t|})~ J(b) ~,
   \label{E12}
\end{equation}
where $J_0(b a \sqrt{|t|})$ is the zeroth--order Bessel function.
Then 
\begin{equation}
T_{H_1 H_2} = i s[\langle g^2 FF\rangle a^4]^2 a^{2} I(t)~. 
\label{4A19A}
\end{equation}
 Since  $J(b)$ is real, $\sigma^T$ and the
slope parameter B are written 
\begin{eqnarray}
            \sigma^T= I_0~[\langle g^2 FF \rangle a^4]^2 a^{2} ~~,~~
   B = \frac {d}{dt} \biggl( \ln \frac {d\sigma^{e\ell}}{dt} \biggr)
   \bigg\vert_{t=0} ~=\frac{1}{2}~ \frac{I_2}{I_0} ~a^2 ~ \equiv K a^2~.
\label {4A21}
\end{eqnarray}

   It is important to observe that these results conveniently factorize
the dimensionless QCD strength $ \langle g^2 FF \rangle a^4$ in the 
expressions for the amplitude, and that the correlation length $a$ 
appears as the natural length scale for the 
observables and for the geometric aspects of the interaction. 
For forward scattering 
 these aspects are concentrated on the quantities $I_0(S_1/a,S_2/a)$ and
$I_2(S_1/a,S_2/a)$ that depend on the hadronic structures and
are  mainly determined by the
values of the profile functions  in the range of impact parameter 
up to about 2.5 fm.  The total cross section 
$\sigma^T$ measures the strength, while the slope B has the strength 
cancelled out and is only related 
to the hadron geometry. 
The explicit formula for the slope is
\begin{equation}
B= \frac{1}{2}\frac{ \int d^2\vec b~b^2~ J(b)}
          {\int d^2\vec b~ J(b)}~a^2= \frac{1}{2}~\langle b^2 \rangle a^2 ~,
\label{4A21A}
\end{equation}
where it is seen as related to the average value of the square of the 
impact parameter in the collision, with J(b) as weight function.
We recall that here b is dimensionless and that  $\langle b^2 \rangle$ 
depends on the hadronic sizes.

  The paper is organized as follows.  In Sec. 2 we present a convenient 
parametrization of the MSV profile function in terms of which the observable 
quantities are calculated.  In Sec. 3 amplitudes and profile functions are 
directly obtained from experimental data and comparison is made between our 
results and those of Carvalho and Menon \cite{MENON1}.  In Sec. 4 we present
general comments and conclusions.
\newpage 

{\bf 2. MSV profile functions} 

  The analytical study  \cite{RRG3A} of the MSV profile functions for 
 large values of b shows that their asymptotic behavior is of the form 
\begin{equation}
 J(b)= \exp(-\rho b) ~ \bigg[\frac{A_1}{b}+
 \frac{A_2}{b^2}+\dots \bigg]~,
\label{E1A}     
\end{equation}
where $A_1, A_2,...$ are functions of $S_1/a$ and $S_2/a$, and the quantity 
$\rho =  3\pi /8$ arises  from the particular form of corrrelation 
function that has been used. For small and intermediate values of $b$  the form of 
$J(b)$ resembles gaussian shapes, which combined with the results of 
MSV for large b leads us to suggest the form
\begin{equation}
  J(b)= [J(0)-a_3] e^{-b^2/a_1} + a_3 e^{-b^2/a_2}+a_4\tilde A_\gamma(b)~ ,
\label{E1}
\end{equation}
with
\begin{equation}
\tilde A_\gamma(b)=\frac{e^{-\rho \sqrt{\gamma^2+b^2}}}{\sqrt{\gamma^2+b^2}}
     \bigg(1-e^{\rho \gamma - \rho \sqrt{\gamma^2+b^2} } \bigg).
\label{E1B}
\end{equation}
The last term fixes the correct behavior of $J(b)$ for large $b$ given 
in  Eq. (\ref{E1A}), with  $\sqrt{\gamma^2+b^2}$  written  
in the place of b, so that the expression  can be extended to b=0.
 The combination inside the parentheses is such that 
$\tilde A_\gamma (0)=0$, and it is shown below that  the use of the two terms 
creates a zero in the imaginary part 
of the amplitude, which is required by the data. 

The parameters $a_1, a_2, a_3, a_4, \gamma $ depend on $S_1$ and $S_2$. 
In pp and $\bar {\rm p}$p scattering we have $S_1=S_2 \equiv S$, and the
behavior of the parameters as functions of $S/a$ is regular and simple.
Good representations are 
\begin{equation}
     J(0)= \frac{0.0024 (S/a)^4}{4.5+(S/a)^3}~ ,
\label{E2}
\end{equation}
\begin{equation}
   a_1= 10 + 0.285 ~  (S/a)^{3}~ ,
\label{E3}
\end{equation}
\begin{equation}
   a_2= 1 + 0.25 ~  (S/a)^{2}~ ,
\label{E4}
\end{equation}
\begin{equation}
 a_3=   \frac{0.002745 (S/a)^3}{4+(S/a)^2} ~,
\label{E5}
\end{equation}
\begin{equation}
 a_4= \frac{4}{2187\pi^2} \bigg[ 1+e^{(\rho S/2a)^2} \bigg]^2 ~ (S/a)^4 ~ , 
  \label{E6}
\end{equation}
and 
\begin{equation}
     \gamma= 0.57~ (S/a)^2 ~ . 
\label{E7}
\end{equation}
The expression for $a_4$ shows the fast increase of the thickness of the  
asymptotic tail as $ S/a$ grows. 
\vspace*{5cm}

     \centerline{ FIGURE 1}

\bigskip

{\small
 Fig. 1~-~ Profile function $J(b)$ of the Model of the Stochastic Vacuum 
for S/a=2.7.  The squares  represent the exact values \cite{RRG3}, and 
the solid line  represents the parametrized $J(b)$ given by Eq. (\ref{E1}) 
with partial  contributions called $G_1$, $G_2$ and $A$.}

The contributions of the terms in 
 Eq. (\ref{E1}) are illustrated in Fig. 1, together with the profile
 $J(b)$ obtained by direct  calculation \cite{RRG3}.
\bigskip

With the analytical expression (\ref{E1}) for $J(b)$, the integrations 
leading to the moments $I_k$  are easily performed, leading to
\begin{equation}
     I_0= 2\pi \bigg( \frac{1}{2}[J(0)-a_3] a_1+\frac{1}{2} a_2 a_3 +
            \frac{a_4}{2 \rho} e^{-\rho \gamma} \bigg) ~ , 
  \label{E8}
\end{equation}
and
\begin{equation}
     I_2= 2\pi \bigg( \frac{1}{2}[J(0)-a_3] a_1^2+\frac{1}{2} a_2^2 a_3 +
            \frac{a_4}{4 \rho^3} e^{-\rho \gamma}
        [ 7 + 6 \gamma \rho ]  \bigg)  ~ . 
  \label{E9}
\end{equation}
 As shown before \cite{RRG3,RRG3A}, these expressions, which are important 
for the evaluation of the observables, can be conveniently  
parametrized as functions of $S/a$, leading to the forms
\begin{equation}
I_0=0.006532~(S/a)^{2.791}
\label{E10A}
\end{equation}
and 
\begin{eqnarray}
  K=\frac{1}{2} \frac{I_2}{I_0}=2.030+0.3293~(S/a)^{2.126} ~ .
\label{E10B}
\end{eqnarray}
  The evaluation of the amplitude through Eqs. (\ref{E12}) and (\ref{4A19A})
requires the integration formula
\begin{equation}
\int_0^\infty J_0(\beta u ) \frac{e^{-\rho \sqrt{\gamma^2+u^2}}}
 {\sqrt{\gamma^2+u^2}} ~ u ~ du = \frac{e^{-\gamma \sqrt{\rho^2+\beta^2}}}
     {\sqrt{\rho^2+\beta^2}} ~ ,
  \label{E13}
\end{equation}
and then we obtain 
\begin{eqnarray}
     I(t)=2\pi \bigg[\frac{1}{2} [J(0)-a_3] a_1 e^{-a^2 |t| a_1/4}
    + \frac{1}{2} a_3 a_2 e^{-a^2 |t| a_2/4}+a_4A_\gamma(t)\bigg]~,
  \label{E14}
\end{eqnarray}
where
\begin{eqnarray}
A_\gamma(t)&=&\int_0^\infty\tilde A_\gamma(b)J_0(ba\sqrt{|t|})bdb 
                                        \nonumber   \\
   &=&  \frac{e^{-\gamma \sqrt{\rho^2+a^2 |t|}}}
   {\sqrt{\rho^2+a^2 |t|}} - e^{\gamma \rho} 
   \frac{e^{-\gamma \sqrt{4 \rho^2+a^2 |t|}}}
   {\sqrt{4 \rho^2+a^2 |t|}}~. 
\label{E14d}
\end{eqnarray}
The difference of two terms in  this expression accounts for a  zero in 
the imaginary amplitude. With the physical value $a=0.32~{\rm fm}$,
the dimensionless combination $a^2 |t|$ takes the reference value 
$a^2 |t|=1 $ for  $|t|= 0.38~ {\rm GeV}^2$.  
For large $|t|$ ( for $ a^2|t| \gg 1$) the largest contributions 
to the amplitude  $I(t)$ come from the last term, with 
\begin{eqnarray}
  I(t)~ \rightarrow ~ 2 \pi ~ a_4 ~ 
      \frac{e^{-\gamma \sqrt{a^2 |t|}}}
   {\sqrt{a^2 |t|}}~ \bigg[
       (1-e^{\gamma \rho})~ 
        -\frac{\gamma \rho^2}{2 \sqrt{a^2 |t|}} (1-4 e^{\gamma \rho})
    +\dots \bigg] ~ ,
  \label{E14a}
\end{eqnarray}
and we observe that for large $|t|$ the amplitude is negative 
($\gamma \rho$ is larger than 1). For small $|t|$ the behavior is 
\begin{eqnarray}
   I(t) \approx2\pi\bigg\{ \frac{1}{2} [J(0)-a_3] a_1 e^{-a^2 |t| a_1/4}
    + \frac{1}{2} a_3 a_2 e^{-a^2 |t| a_2/4}+
a_4{e^{-\rho\gamma}\over{2\rho}}\times \nonumber \\
\bigg[1-\bigg({3\gamma\over4\rho}+
 {7\over 8\rho^2}\bigg)a^2|t|+...\bigg]\bigg\}~,~~~~~~~~~~~~~~~~~~
  \label{E14b}
\end{eqnarray}
where the last two terms are small compared to the first, so that the 
amplitude has the typical exponential behavior.  However, we remark that the 
amplitude is not of the simple form $ T(s,|t|)=T(s,0) e^{-B|t|/2}$, 
and the slope $B$  must be evaluated through  Eqs. (\ref{4A21}), 
(\ref{E8}), (\ref{E9}). For t=0 we have  $I_0=I(0)$ consistently with 
Eq. (\ref{E8}).

The results obtained with MSV are correctly related to experimental 
data at t=0 (total cross section and slope), leading to the determination 
of the QCD parameters \cite{RRG3,RRG3A}.
However, the predictions for the differential cross sections beyond the
small values of $|t|$ are not correct, mainly because the MSV profiles 
put the zeros of the imaginary amplitudes at too large values of $|t|$. 
In the next section we introduce empirical modifications in the parameters 
in order to reproduce the  data in the full  $|t|$ range.
\bigskip

{\bf 3. Amplitudes and profile functions from the experimental data}

  Using the same form of parametrization suggested by MSV, we construct the 
dimensionless amplitude 
\begin{equation} 
T_{\rm exp}(s,t)=4\sqrt{\pi}s[i{\cal I}(t)+{\cal R}(t)]
\label{E36}
\end{equation}
directly from the experimental data (hereafter, expressions such as 
experimental 
amplitude, experimental profile and so on will be used for quantities 
obtained directly from the data). 
The imaginary and real  parts of the amplitude are 
parametrized using the forms
\begin{equation}
{\cal I}(t)=  \alpha_1e^{-\beta_1|t|}+\alpha_2e^{-\beta_2|t|}+
\lambda~2\rho e^{\rho\gamma}A_\gamma(t)
\label{E15}
\end{equation}
and        
\begin{equation}
{\cal R}(t) =  \lambda^\prime~
2\rho e^{\rho\gamma^\prime}A_{\gamma^\prime}(t) 
      +\alpha^{\prime}_1e^{-\beta^{\prime}_1|t|}~,
\label{E16}
\end{equation}
where $\rho=3 \pi /8$ and where we have grouped factors 
$2\rho e^{\rho\gamma}A_\gamma(t)$ 
in order to have $2\rho e^{\rho\gamma}A_\gamma(0)=1$. 
We use 8 parameters for each energy (here we do not count 
$\alpha^\prime_1$ and $\beta^\prime_1$ which are the same for all 
energies, as explained below).

The use of $A_\gamma(t)$ in the real part, which is inspired in the 
MSV form for the imaginary part,  has a more convenient structure, 
compared to simple exponentials, to fill the dip left by the zero of 
the imaginary part. On  the other hand, the simple exponential term 
$\alpha^\prime_1~{\rm exp}(-\beta^\prime_1\vert t\vert)$ 
was included in the real part specifically to describe the large 
$\vert t\vert$ ($5<\vert t\vert<15~\rm{GeV}^2$) data \cite{RRDAT5}  
at 27 GeV, which  is assumed to be universal for all ISR energies. 
This term is not present in our description of the data at 
546 and 1800 GeV, where large $|t|$ values have not been measured.

  We use as experimental inputs for pp scattering at ISR energies and 
for $\rm{\bar p p}$ at 
546 and 1800 GeV the ratio of the forward real to imaginary parts of 
the amplitude, ${\cal R}(0)/{\cal I}(0)\equiv\bar\rho$, the slope at $t=0$, 
\begin{equation}
B={d\over dt}\biggl(\ln {d\sigma\over dt}\biggr)\bigg\vert_{t=0}=
          2{{{\cal R}(0){\cal R}^\prime(0)+
{\cal I}(0){\cal I}^\prime(0)}\over{{\cal R}^2(0)+{\cal I}^2(0)}} ~ ,
\label{E18}
\end{equation}
and the total cross section determined through the optical theorem, 
\begin{equation}
\sigma^T=4\sqrt{\pi}{\cal I}(0)~.
\label{E18a}
\end{equation}
 After this information is introduced, the remaining parameters are 
determined by fitting 
$|{\cal R}(t)+i{\cal I}(t)|^2$ to the experimental data for $d\sigma/dt$. 
The results of fittings at the ISR energies and at 546 and 1800 GeV are 
shown in  Figs. 2 and 3.  


\vspace*{5cm}
{\small
 Fig. 2 ~-~Results of fittings of elastic differential cross sections 
through Eqs. (\ref{E15}) and (\ref{E16}) 
at ISR energies. Curves and data are  conveniently separated in the 
figure through multiplication by the factors that are shown in the 
upper corner. The large $|t|$ data measured at 27 GeV that are drawn 
together with all angular distributions are described by the last term 
of the real amplitude in Eq.(\ref{E16}). }

\vspace*{5cm}

        \centerline{ FIGURE 3 }
{\small
 Fig. 3 ~-~Same as in Fig. 2 , for 546 GeV and 1800 GeV.}
\bigskip

  The parameters present a very regular energy dependence, as shown in  
Table II and Fig. 4. However, it is dificult to write parametrized 
analytical dependences for  the parameters as functions of $\sqrt{s}$ 
because the expressions given by Eqs. (\ref{E15}) and (\ref{E16}) are 
very sensitive to the values of the parameters, requiring accuracy of 
at least three decimals to reproduce correctly the data if we require 
smoothness in the variation with the energy. If each energy is 
considered separately, and allowing for the experimental errors in
the data, the parameters are fixed typically within 1\% at the 
energies for which the data are more accurate (e.g. 52.8 GeV).

\vspace*{5cm} 
\centerline{Table  II}

Table II - Values of the parameters for Eqs. (\ref{E15}) and (\ref{E16}). 

\bigskip

\vspace*{5cm}

  \centerline{ FIGURE 4} 

{\small
 Fig. 4~-~ Parameters of the fittings through Eqs. (\ref{E15}) and 
(\ref{E16}) at ISR energies, and at 546 and 1800 GeV.  The energy 
dependence of the total cross section $\sigma^T$ 
and the slope $B$ obtained from the parameters are also shown.}
\bigskip

The imaginary and real parts of the experimental scattering amplitude 
$T_{\rm exp}(s,t)$ are shown in Figs. 5 and 6 respectively. The figures 
show remarkable regularities in the energy dependence, with an interesting
similarity in the behavior of all curves as functions of $|t|$. The 
real parts grow with the energy faster than the imaginary parts, so 
that even after multiplication  by 1/s the real parts still have 
magnitudes which increase  with s (note the inverted ordering of the 
curves in   the two figures). At all investigated energies the 
imaginary amplitudes present only one zero as functions of $|t|$, 
 with a regular displacement of the zeros towards the origin as the 
energy increases (shrinking of the diffraction peak). The prediction 
of a second zero in the imaginary part by Borrely, Soffer and
Wu  \cite{WU} for very high energies ( above 10 TeV) is very far 
for the range of our analysis.
 Depending on the sign given to the last term in Eq. (\ref{E16}), the real 
amplitude may present one (for negative sign) or two zeros (case of 
positive sign). This last term practically does  not 
interfere with the first part, as its contribution is only important
in the tail of the distribution, and thus the fitting does not
determine its sign. The 
first zeros of the real parts are distributed in  a rather small range 
of $|t|$ values, closer to the origin compared to the zeros of the 
imaginary parts. The displacement of the zeros  with the energy is shown 
in Fig. 7 (only the first zeros of the real parts are shown).  

\vspace*{5cm}

       \centerline{FIGURE 5}  

{\small
 Fig. 5 ~-~ Imaginary part of the experimental scattering amplitude 
at the ISR energies and at 546 and 1800 GeV, divided by s.} 
\vspace*{5cm}

     \centerline{FIGURE 6} 

{\small
 Fig. 6~-~ Same as in Fig. 5, but for the real part of the amplitude.}
\bigskip

\vspace*{5cm}

\centerline{FIGURE 7}

{\small
 Fig. 7~-~ Locations of the zeros of the imaginary and real parts 
    of the amplitude as functions of the energy. }
\bigskip

Unfortunately the data for $d\sigma/dt$  at 1800 GeV do not go 
beyond $|t|=0.6~ {\rm GeV}^2$, and the positions of the zeros 
and the large $|t|$ behavior cannot be extracted from the data. 
Actually, Figs. 5, 6, 7 only give (extrapolated) guesses for the 
quantities related to this energy.

Data at large $|t|$ (up to  15 GeV$^2$) obtained \cite{RRDAT5}  
only at 27 GeV seem to offer a natural extrapolation for the angular 
distribution at all ISR energies, as shown in Fig. 2. This tail
is described by the term 
$ \alpha^{\prime}_1e^{-\beta^{\prime}_1|t|}$ in the real part
in Eq. (\ref{E16}), 
with small values for $ \alpha^{\prime}_1$ and  $\beta^{\prime}_1$, 
which hardly influences  the determination of the other parameters. Thus 
we have a term specific for the energy independent tail, which has
been included in the curves of Fig. 2.
 The large $|t|$ tail in $d\sigma/dt$ has been parametrized \cite{RRB2a}
in the form $0.09/t^8$, and to  have an expression with this behavior and
 that at the same time can be used for all $|t|$, we may  
modify Eq. (\ref{E16}) to have  the form
\begin{equation}
{\cal R}(t) =  \lambda^\prime~
2\rho e^{\rho\gamma^\prime}A_{\gamma^\prime}(t)+ 
  \alpha^{\prime\prime}_1[1-e^{-\beta^{\prime\prime}_1/(a^2|t|)^4}]
 ~.
\label{E16a}
\end{equation}
We discuss this representation in the next section.

  To obtain the profile functions from the experimental data, we 
invert the Fourier-Bessel transform
\begin{equation}
{\cal R}(t)+i{\cal I}(t)=2\pi\int_{0}^{\infty}[{\cal K}(\tilde b)+
     i{\cal J}(\tilde b)]
J_0(\tilde b\sqrt{|t|})\tilde bd\tilde b ~ ,
\label{E22}
\end{equation}
where $\tilde b\equiv ab$ (here $\tilde b$ is in fermi or 
${\rm GeV}^{-1}$ units).  We then obtain the desired 
(dimensionless) imaginary and real parts of the profile function 
\begin{equation}
{\cal J}(\tilde b)={1\over 4\pi}\biggl[
 {\alpha_1\over \beta_1}{\rm e}^{-{\tilde b}^2 /4\beta_1}+
 {\alpha_2\over \beta_2}{\rm e}^{-{\tilde b}^2 /4\beta_2}+
4{\lambda\rho\over a^2} {\rm e}^{\rho\gamma}{\tilde A}_\gamma (\tilde b/a)\biggr]
\label{E23}
\end{equation}
and
\begin{equation}
{\cal K}(\tilde b)={1\over 4\pi}\biggl[
4{\lambda^\prime\rho\over a^2} {\rm e}^{\rho\gamma^\prime}
{\tilde A}_{\gamma^\prime} (\tilde b/a)+
 {\alpha^\prime_1\over \beta^\prime_1}{\rm e}^{-{\tilde b}^2 /4\beta^\prime_1}
       \biggr] ~ .
\label{E23a}
\end{equation}

  To follow  the same notation used for the profile function of MSV (see 
Eq. (\ref{E1}) and recall that in MSV we have only the imaginary part 
of the amplitude) we must introduce a numerical factor 
\begin{equation} 
J_{\rm exp}(\tilde b)={4\sqrt{\pi}\over [\langle g^2FF\rangle a^4]^2}
{\cal J}(\tilde b)~ ,
\label{E28}
\end{equation}
with $4\sqrt{\pi}/[\langle g^2FF\rangle a^4]^2 =2.02\times10^{-2}~ .$

  Given the experimental profiles ${\cal K}(\tilde b)$ and 
${\cal J}(\tilde b)$, we can 
now write convenient expressions for comparison between the experimental 
quantities and those of MSV. We define the moments of the experimental 
profile functions corresponding to the imaginary  part of the scattering 
amplitude, 
\begin{equation}
{\cal I}_k \equiv 2\pi\int_{0}^{\infty}\tilde{b}^k{\cal J}
      (\tilde b)\tilde bd\tilde b~,
\label{E24} 
\end{equation}
and to the real part
\begin{equation}
{\cal R}_k \equiv 2\pi\int_{0}^{\infty}\tilde{b}^k{\cal K}
           (\tilde b)\tilde bd\tilde b~,
\label{E24a} 
\end{equation}
which depend only on $\sqrt{s}$.  The moments can be evaluated analytically,
and are given by expressions similar to Eqs. (\ref{E8}) and (\ref{E9}).    

   Including the real part, the expression for the slope 
parameter at $t=0$ in terms of the moments ${\cal R}_k$ and ${\cal I}_k$ is 
\begin{equation}
B_{\rm exp}={{\cal R}_0{\cal R}_2+{\cal I}_0{\cal I}_2\over
2({\cal R}_0^2+{\cal I}_0^2)}=
   {\bar\rho{\cal R}_2+{\cal I}_2\over
   2(1+\bar\rho^2){\cal I}_0} ~ ,
 \label{E27b}
\end{equation}
where  ${\cal R}_0/{\cal I}_0=\bar\rho$ . 
The maximum error in $B_{\rm exp}$ caused by the suppression of  
the real part contribution is about 2\% at 1800 GeV and less than 
1\% at the ISR energies. 

In order to compare experimental quantities with those of MSV, we must 
have a correspondence between the experimental values of $\sqrt{s}$ and 
the MSV values of $S/a$ for the proton radius \cite{RRG3,RRG3A}. We then 
require that
\begin{equation}
I_0(S/a)={\cal I}_0(s){4\sqrt{\pi}\over [\langle g^2FF\rangle a^4]^2 a^2}~,
\label{E37}
\end{equation}
which means that the evaluated total cross sections must be the same.
The solutions of this  equation can be written in a simple parametrized 
form 
\begin{equation}
S/a =~2.1848~+~0.1623~\ln{\sqrt{s}}~.
\label{E37a}
\end{equation}
 According to the phenomenology of the application of MSV to pp and 
$\bar {\rm p}$p scattering \cite{RRG3,RRG3A}, the experimental and MSV values 
of $\sigma^T$ and $B$ are fixed to be the same for each one of the two 
input energies (541, 1800 GeV). This fixes
the two fundamental QCD quantities (gluon condensate  and correlation 
length)  and the values of the effective proton 
radius at each of these two energies. Then 
  the MSV and experimental calculations of $I_2$ (and obviously the 
ratio $I_2/I_0$ that gives the slope $B$) agree within 1 to 2\% accuracy at all
energies. This is the basis of the success of the MSV calculation 
of total cross sections  and slopes.

  It is important to compare our results with those 
obtained by P. A. Carvalho and M. J. Menon (CM) \cite{MENON1}, who 
parametrized the scattering amplitude (here rewritten according to 
our notation) using  a sum of exponentials in the form 
\begin{equation}
F(s,t)=~-\mu(s)\sum_{j=1}^2 \alpha_j {\rm e}^{-\beta_j \vert t \vert}+
      i~\sum_{j=1}^n \alpha_j {\rm e}^{-\beta_j \vert t \vert}~,
\label{ME1}
\end{equation}
where 
\begin{equation}
\mu (s)=~ -{\bar \rho (s) \over (\alpha_1+\alpha_2)} \sum_{j=1}^{n} \alpha_j~,
\label{ME2}
\end{equation}
with coefficients obtained by fitting $d\sigma/dt=\pi \vert F(s,t)\vert^2$ 
to the pp experimental data at the ISR energies (n= 4 - 6, depending on the
energy).  The corresponding 
dimensionless profile functions are
\begin{equation}
{\cal K}_{\rm CM}(\tilde b)=-{\mu(s)\over 2}\sum_{j=1}^2 {\alpha_j \over\beta_j}
{\rm e}^{-{\tilde b}^2 /4\beta_j}
\end{equation}
for the real part, and
\begin{equation}
{\cal J}_{\rm CM}(\tilde b)={\frac{1}{2}} \sum_{j=1}^n {\alpha_j\over\beta_j}
{\rm e}^{-{\tilde b}^2 /4\beta_j}~
\label{ME4}
\end{equation}
for the imaginary part. Comparison with our formulas is given by the 
correspondence of values
\begin{eqnarray}
{\cal K}_{\rm CM}(\tilde b)= ~2\sqrt{\pi}{\cal K}(\tilde b)~ ~, ~ ~
{\cal J}_{\rm CM}(\tilde b)= ~2\sqrt{\pi}{\cal J}(\tilde b)~.
\label{ME5}
\end{eqnarray}

 Our imaginary amplitudes 
  $~{\rm Im}[T_{\rm exp}(s,t)]/4\sqrt{\pi}s={\cal I}(t)~$ 
and the values of $\sqrt{\pi}{\rm Im}[F(s,t)]$ obtained by Carvalho and Menon 
are very similar  for all ISR energies. The Fourier tranforms
${\cal J}(\tilde b)$ and ${\cal J}_{\rm CM}(\tilde b)$, however,  
coincide  only between b=0 and b=2.  
 On the other hand the real amplitudes 
${\rm Re}[T_{\rm exp}(s,t)]/4\sqrt{\pi}s={\cal R}(t)$ 
and $\sqrt{\pi}{\rm Re}[F(s,t)]$  differ substantially for all values 
of $|t|$, except at 
$|t|=0$ where they coincide (within the experimental error), 
because of the input experimental value for $\bar\rho$. The comparison 
of the real parts obtained in the two works is illustrated in Fig. 8 for 
the amplitudes  and in Fig. 9 for the profile functions.

\vspace*{5cm}

       \centerline{FIGURE 8} 
\bigskip

{\small
 Fig. 8  ~-~ Comparison of the real parts of the experimental amplitudes 
at 23.5 GeV obtained in our work (solid curve) and by Carvalho and 
Menon (CM) \cite{MENON1} (dashed curve). } 
\vspace*{5cm}

    \centerline{ FIGURE 9} 

{\small
 Fig. 9 ~-~ Same as Fig. 8, but for the real parts of profile functions, with 
the correspondence between profiles given in Eq. (\ref{ME5}).
In our representation (solid line) the value of the profile function
at $b=0$ is fully given by the second term of  the amplitude, that only 
refers to the tail. }
\bigskip

In spite of these differences  in the real parts, the two 
parametrizations are equivalent with respect to the description of the data
at each separate energy. However, our results are distinguished by the
very regular energy dependence of the parameters. This regularity, which is
important for interpolations and extrapolations, is true up to the 
highest energies 540  and 1800 GeV and also includes the angular 
distribution  at large $|t|$ measured at 27 GeV.    

\bigskip

 {\bf 4. Comments and Conclusions}

  The application of MSV to high-energy scattering leads to a good 
description of the observables determined by the forward amplitude
($\sigma^T$ and $B$), but  the predictions for the  differential 
cross sections for large  $|t|$ are not correct, mainly because the 
MSV profiles put the zero of the imaginary amplitude at values of $|t|$ 
that are systematically  higher than those required by the data. The 
model  gives a dynamical, QCD based, framework to evaluate the forward
amplitude in the eikonal formalism, and  in its present formulation 
does not intend to describe elastic scattering at all $|t|$, that  would 
require the inclusion of a real part in the amplitude and changes 
in the imaginary part.  In the present work,  starting  from the 
analytical forms of profile function and amplitude suggested by the model, 
we investigate  the modifications needed to fit all elastic scattering 
data of  the pp and 
$\bar{\rm p}{\rm p}$ systems at high energies. 
 Values are obtained for the parameters and we believe  that the 
results are made rather unique through the effort to 
obtain  a smooth energy dependence.
The  chosen form respects the properties of the profile function and 
is such that the 2-dimensional Fourier transform 
can be written analytically. The existence of a zero in the imaginary 
amplitude is a common feature of both the MSV calculation  and the 
experimental data, and the  measured differential 
cross section requires that this zero be filled up with the contribution
from the real part. We thus must add a real part to the MSV amplitude. 

 In our parametrization  the 27 GeV tail  
($5\gappl |t| \gappl 15 ~ {\rm GeV}^2$) is specifically 
 described by a term in the real part. 
The large $|t|$ amplitude is probably   dominated  by perturbative 
contributions, such as triple gluon exchange, which is real and can 
account for the observed $ 0.09/t^8 $ behavior \cite{RRB2a}, although 
it is surprising that there is no energy dependence.   
Fig. 10  shows the comparison between the exponential form used in 
Eq. (\ref{E16}) and the power behavior  produced by Eq. (\ref{E16a}). 
In this last case the values of the parameters are 
$ \alpha^{\prime \prime}_1= 0.001~ {\rm GeV}^{-2}$ and 
$\beta ^{\prime \prime}_1= 23650$. Both forms describe well the data.

\vspace*{5cm}
    
          \centerline{FIGURE 10} 

{\small
 Fig. 10 ~-~ Differential cross section for large $|t|$ values.  
 Comparison between the parametrization given by Eqs. (\ref{E15}) and 
 (\ref{E16}), and the  $1/|t|^8$ dependence \cite{RRB2a} produced 
by Eq. (\ref{E16a}).} 
\bigskip

    In Table II we have chosen a positive sign for the tail term 
added to the real part, although it could also be negative, the 
different choices only requiring a slight adjustment of 
parameters. A positive sign in this term causes a second zero 
in the real amplitude occurring at $|t| \approx 2$ GeV$^2$, but there
appears no   dip because the imaginary part is large in this region.
The first zero of the real amplitude occurs at very small $|t|$
for either sign of the tail term. The roles of the real and imaginary
parts of the amplitude, and separately of the real term for the 
tail, considering its two possible signs, are shown in Fig. 11. 

\vspace*{5cm}

    \centerline{ FIGURE 11}

{\small
 Fig. 11 ~-~ Separate roles of $|{\rm Im} ~ T|^2$ and $|{\rm Re}~ T|^2$ 
at 23.5 GeV.
The dip in the data at about 1.5 GeV$^2$ is due to a zero in ${\rm Im}~T$. 
The real part is positive at $t=0$, becomes negative at small 
$|t|$ and fills partially the dip in $d\sigma/dt$. The dotted line 
represents the squared tail term alone. If this term is positive, 
it causes a second zero when added to ${\rm Re} ~ T$, as shown by the (+) 
line. In the (-) line  the tail is negative and  no second zero exists. }
\bigskip

 Fig. 12 shows the profile functions obtained from MSV and from the 
fitting to the data.  A general fact is that the experimental profile 
functions are lower for small $b$ and higher for intermediate $b$ values, 
compared to the MSV shapes.  Due to the product $\tilde b J(\tilde b)$ 
that appears in the 2-dimensional Fourier transform, the region about 
$\tilde b\simeq0$ has little physical significance.  It is then more 
instructive to examine the forms of the product $\tilde b J(\tilde b)$, 
as shown in Fig. 13. We remark that in this figure corresponding MSV and 
experimental profiles have the same area (the moments  $I_0 $ fix the total 
cross section). In a plot of $b^3 J(b)$ the corresponding profiles
also have the same areas, since the MSV moments $I_2 $ yield the 
correct slopes. 

\vspace*{5cm} 
 
    \centerline{ FIGURE 12}  

{\small
 Fig. 12  ~-~ Comparison of values of $J(b)$ from experimental data and 
 from MSV, for corresponding values of $\sqrt{s}$ and $S/a$ according to 
 the requirement established by Eq. (\ref{E37}).}
\bigskip

\vspace*{5cm}

     {\centerline{ FIGURE 13}

{\small
 Fig. 13~-~ Same as in Fig. 12, but for $bJ(b)$.}
\bigskip
 
 The present development of the MSV  formalism applied to 
 hadron-hadron scattering \cite{RRG3} is restricted to the lowest order 
nonvanishing contribution, which is quadratic in the gluonic correlator, 
 and has been shown to be sufficient 
 for the evaluation of total cross sections and slope parameters.  
The resulting amplitude is purely imaginary, and the ratio  of the real to the 
imaginary parts of the elastic scatering amplitude can only be described if 
we go one further order in the contributions of the corrrelator. 
On the other hand, we have verified that changes in the form  of the 
correlation functions $D(z^2/a^2)$ and $D_1(z^2/a^2)$, that enters the 
definition of the correlator, and reasonable changes  of the hadron 
wave-function, have little effect on the results.

  In addition to these considerations, which concern  mainly  the 
nonperturbative nature of the model, we must also take   
perturbative contributions in account. To the MSV calculation, which 
represents single pomeron exchange, we may add  terms with mixed 
pomeron and multiple gluon exchanges, which  modify the imaginary 
and  real parts, and may add three  gluon exchange (purely real) for 
the tail \cite{RRB2a}.

In Fig. 14  we plot the difference at 23.5  GeV between 
the experimental and MSV imaginary parts of the  amplitude $T(s,t)$
$$ {1\over s}~ [{\rm Im}~T_{\rm exp}(s,t)-{\rm Im}~T_{\rm MSV}(s,t)]~,$$ 
which illustrates the contribution  to be added to  the MSV result for 
the imaginary part, which must be mostly negative, peaked at about 
$  |t|= 0.5~ {\rm GeV}^2 $. In the same figure  we plot the real part  that 
is  extracted  from the data. We observe that, except for a small
range of very small $|t|$ values, the  corrections to the imaginary part 
must be much larger than the values of the real amplitude. 
The situation is similar for the other energies.
These remarks may guide future theoretical developments.

\vspace*{5cm}

  \centerline {FIGURE 14} 

{\small
 Fig. 14 ~-~ Real part of the amplitude and difference between the 
imaginary parts of the amplitude obtained from the experiments and from 
the MSV calculation, all at 23.5 GeV.  }

\bigskip

 Fig. 15 shows the real parts of the profile function for all energies 
of our investigation. We observe a very regular energy dependence. 
The quantity $ {\tilde A}_{\gamma^\prime} (\tilde b/a)$
 in Eq. (\ref{E23a}) is zero for b=0, so that the value
at the origin is due only to the tail term, being the same for all
ISR energies, and not included at 540 and 1800 GeV.

\vspace*{5cm} 

   \centerline{FIGURE 15}

{\small
  Fig. 15~-~ Real part  ${\cal K}(b)$ of the experimental profile 
functions for all ISR energies, 546 GeV and 1800 GeV.}
\bigskip

   Now we comment on the relation between observables in high-energy 
scattering and the hadronic sizes. 
  The geometrically extended character of the non-perturbative QCD
interaction determines the phenomenological properties of the observables, 
which are fixed by the sizes and global structures of the colliding 
systems. These features have led to models of 
geometric nature for high-energy scattering \cite{RRG3D},
that give a natural account of the relations between total cross sections 
and slopes of the diffractive peaks of different hadronic systems. 
  A geometrical relation of this sort is seen if we write 
Eq. (\ref{4A21A}) in the form
\begin{equation}
B=~\frac{1}{2} \frac{\int d\tilde b {\tilde b}^2[\tilde b J(\tilde b)]}
   {\int d\tilde b [\tilde b J(\tilde b)]}=
   \frac{1}{2}\langle {\tilde b}^2\rangle~.
\label{4A21Aa}
\end{equation}
and  interpret $\langle {\tilde b}^2 \rangle$ as the average value 
of the square of the impact parameter taking  $\tilde b J(\tilde b)$ as 
weight function.  There is  an interesting  relation 
between this average value $\langle {\tilde b}^2\rangle$ and 
the square of the value ${\tilde b}^{(3)}_{\rm m}$
of the impact parameter at the peak  of the function 
${\tilde b}^3 J(\tilde b)$ that appears in  the upper integrand of 
Eq. (\ref{4A21Aa}).  We have verified that the ratio 
$[{\tilde b}^{(3)}_{\rm m}]^2/\langle {\tilde b}^2\rangle$ 
of these two quantities extracted from the data is a 
constant in a wide energy range, as shown in Fig. 16.
In the case of a pure gaussian profile (i.e. an amplitude
with  a single exponential in t) this ratio is 1.5 .
 This property help  us to  define  an effective  hadronic radius,
related to ${\tilde b}^{(3)}_{\rm m}$, in terms of a measurable  quantity.  
It is remarkable, and may be of practical importance for the 
determination of slopes, that the position  
${\tilde b}^{(3)}_{\rm m}$ of the peak is not dramatically  sensitive to 
the details of the amplitude close to  $ t=0$.

\vspace*{5cm}

\centerline {FIGURE 16} 

{\small
 Fig. 16~-~ Values of the experimental ratio 
$[b^{(3)}_{\rm max}]^2/\langle b^2\rangle$
 (black squares) and its mean value (line).} 
\bigskip
 
A convenient parametrization of the experimental values of 
$\tilde b_{\rm max}^{(3)}$ is 
\begin{equation}
\tilde b_{\rm max}^{(3)}=1.0119+0.0433~{\rm ln}{\sqrt{s}}~,
\label{E29}
\end{equation}
and with the correspondence between $S/a$ and $\sqrt{s}$ given by 
Eq. (\ref{E37a}), we may also write
\begin{equation}
\tilde b_{\rm max}^{(3)}=0.4290+0.2668~(S/a)~.
\label{E30}
\end{equation}

The quantity $\tilde b_{\rm max}^{(3)}$ (written above in fermis, while 
$\sqrt{s}$ is in GeV) can be extracted from the amplitude  and gives a 
measurement of the effective hadronic size. 
Eq. (\ref{E30}) connects the effective hadronic collision radius 
with the radius of the transverse wave-function. As the energy 
varies from 23 to 1800 GeV,   $\tilde b_{\rm max}^{(3)}$ varies from 
1.15 to 1.34 fm. We obtain for $\sqrt{s}$=14 TeV the predictions 
$\tilde b^{(3)}_{\rm max}=1.43$ fm, $S/a=3.73$, and 
$B_{\rm exp}=18.76~ {\rm GeV}^{-2}$. 

\bigskip

{\bf Acknowledgements }

   The authors  wish to  thank M. J. Menon, P. A. Carvalho and A. Martini
 for information on their work and for numerical data of experimental 
measurements.



\centerline{ TABLES}
\bigskip

\begin{center}
\begin{table}
\caption{Experimental Data}
\vspace{.5 cm}
\begin{tabular}{|c|c|c|c|c|c|c|}
\hline
& & & & & & \\
& $ \sqrt{s} $ & $\sigma^T $ & $B $ & $ {\rm Ref.}$&$\bar\rho$&${\rm Ref.}$\\
& $(\rm{GeV})$ & $(\rm{mb})$ & $(\rm{GeV}^2)$ & \cite{RRDAT1} &$ $&\cite{RRDAT1}\\
\hline
&     $23.5$&$39.65\pm0.22$&$11.80\pm 0.30$&$(a)$&$0.020\pm0.050$&$(b,c)$\\
&     $30.7$&$40.11\pm0.17$&$12.20\pm 0.30$&$(a)$&$0.042\pm0.011$&$(b,c)$\\
 pp & $44.7$&$ 41.79\pm0.16$&$12.80\pm 0.20$&$(b)$&$0.062\pm0.011$&$(b,c)$\\
&     $52.8$&$42.38\pm0.15$&$12.87\pm 0.14$&$(a)$&$0.078\pm0.010$&$(b,c)$\\
&     $62.5$&$43.55\pm0.31$&$13.02\pm 0.27$&$(a)$&$0.095\pm0.011$&$(b,c)$\\
\hline
$\bar{\rm p}$p   & $ 541$&$62.20\pm 1.50$&$15.52\pm 0.07$&$(d)$ & $0.135\pm
0.015$ & $(d)$\\
&    $ 1800$&$72.20\pm 2.70$&$16.72\pm 0.44$&$(e)$ & $0.140\pm0.069$ & $(e)$\\
\hline
\end{tabular}
\end{table}
\end{center}

\bigskip
%

%
\begin{center}
\begin{table}
\caption{Values  of parameters for Eqs. (\ref{E15}) and (\ref{E16}) }
\vspace{.5 cm}
\begin{tabular}{|c|c|c|c|c|c|c|c|c|c|}
\hline
 & & & & & & & & &\\
$\sqrt{s}$&$\alpha_1$&$\beta_1$&$\alpha_2$&$\beta_2$&$\lambda 
$&$\lambda^\prime$&$\gamma $&$\gamma^\prime$&$\chi^2$\\
 & & & & & & & & &\\
\hline
$~~23.5~$&$1.7298$&$1.4390$&$3.1091$&$2.2949$&$~9.1850$&$0.2774$&$3.8620$&$6.00$&$
1.16$\\
$~~30.7~$&$1.8224$&$1.4502$&$3.1649$&$2.3299$&$~9.5467$&$0.6073$&$3.9000$&$6.600$&
$1.21$\\
$~~44.7~$&$1.6699$&$1.5025$&$3.0000$&$2.1086$&$10.3650$&$0.9354$&$4.0311$&$6.550$&
$4.28$\\
$~~52.8~$&$1.8500$&$1.5287$&$2.9600$&$2.1753$&$10.4630$&$1.1913$&$4.0540$&$6.685$&
$3.22$\\
$~~62.5~$&$1.9272$&$1.5529$&$2.8081$&$2.1337$&$10.8201$&$1.4747$&$4.0924$&$7.300$&
$1.75$\\
$~546~  
$&$2.6174$&$2.1539$&$3.9061$&$2.1813$&$15.7463$&$3.0270$&$4.8190$&$7.970$&$1.23$\\
$1800~  
$&$3.1036$&$2.4526$&$4.4246$&$2.4253$&$18.3315$&$3.6204$&$5.3450$&$8.600$&$2.00$\\
\hline
\multicolumn{10}{|c|}{  }\\
\multicolumn{10}{|c|}{ $\alpha_1$, $\beta_1$, $\alpha_2$, 
           $\beta_2$, $\lambda$ and $\lambda^\prime$ are in $\rm{GeV}^{-2}.~ ~ $
      ~ ~  $\gamma$, $\gamma^\prime$ are dimensionless.}\\ 
\multicolumn{10}{|c|}{$\alpha^\prime_1=0.0031~\rm{GeV}^{-2}$ and 
    $\beta^\prime_1=0.41~\rm{GeV}^{-2}$ are the same for all ISR energies.}\\
\multicolumn{10}{|c|}{  }\\ 
\hline
\end{tabular}
\end{table}
\end{center} 

\newpage

\centerline { FIGURE CAPTIONS}

\bigskip

{\small
 Fig. 1~-~ Profile function $J(b)$ of the Model of the Stochastic Vacuum 
for S/a=2.7.  The squares  represent the exact values \cite{RRG3}, and 
the solid line  represents the parametrized $J(b)$ given by Eq. (\ref{E1}) 
with partial  contributions called $G_1$, $G_2$ and $A$.}

\bigskip

{\small
 Fig. 2 ~-~Results of fittings of elastic differential cross sections 
through Eqs. (\ref{E15}) and (\ref{E16}) 
at ISR energies. Curves and data are  conveniently separated in the 
figure through multiplication by the factors that are shown in the 
upper corner. The large $|t|$ data measured at 27 GeV that are drawn 
together with all angular distributions are described by the last term 
of the real amplitude in Eq.(\ref{E16}). }

\bigskip

{\small
 Fig. 3 ~-~Same as in Fig. 2 , for 546 GeV and 1800 GeV.}

\bigskip

{\small
 Fig. 4~-~ Parameters of the fittings through Eqs. (\ref{E15}) and 
(\ref{E16}) at ISR energies, and at 546 and 1800 GeV.  The energy 
dependence of the total cross section $\sigma^T$ 
and the slope $B$ obtained from the parameters are also shown.}

\bigskip

{\small
 Fig. 5 ~-~ Imaginary part of the experimental scattering amplitude 
at the ISR energies and at 546 and 1800 GeV, divided by s.} 

\bigskip

{\small
 Fig. 6~-~ Same as in Fig. 5, but for the real part of the amplitude.}

\bigskip

{\small
 Fig. 7~-~ Locations of the zeros of the imaginary and real parts 
    of the amplitude as functions of the energy. }

\bigskip

{\small
 Fig. 8  ~-~ Comparison of the real parts of the experimental amplitudes 
at 23.5 GeV obtained in our work (solid curve) and by Carvalho and 
Menon (CM) \cite{MENON1} (dashed curve). } 

\bigskip

{\small
 Fig. 9 ~-~ Same as Fig. 8, but for the real parts of profile functions, with 
the correspondence between profiles given in Eq. (\ref{ME5}).
In our representation (solid line) the value of the profile function
at $b=0$ is fully given by the second term of  the amplitude, that only 
refers to the tail. }

\bigskip

{\small
 Fig. 10 ~-~ Differential cross section for large $|t|$ values.  
 Comparison between the parametrization given by Eqs. (\ref{E15}) and 
 (\ref{E16}), and the  $1/|t|^8$ dependence \cite{RRB2a} produced 
by Eq. (\ref{E16a}).} 

\bigskip

{\small
 Fig. 11 ~-~ Separate roles of $|{\rm Im} ~ T|^2$ and $|{\rm Re}~ T|^2$ 
at 23.5 GeV.
The dip in the data at about 1.5 GeV$^2$ is due to a zero in ${\rm Im}~T$. 
The real part is positive at $t=0$, becomes negative at small 
$|t|$ and fills partially the dip in $d\sigma/dt$. The dotted line 
represents the squared tail term alone. If this term is positive, 
it causes a second zero when added to ${\rm Re} ~ T$, as shown by the (+) 
line. In the (-) line  the tail is negative and  no second zero exists. }

\bigskip

{\small
 Fig. 12  ~-~ Comparison of values of $J(b)$ from experimental data and 
 from MSV, for corresponding values of $\sqrt{s}$ and $S/a$ according to 
 the requirement established by Eq. (\ref{E37}).}

\bigskip

{\small
 Fig. 13~-~ Same as in Fig. 12, but for $bJ(b)$.}

\bigskip

{\small
 Fig. 14 ~-~ Real part of the amplitude and difference between the 
imaginary parts of the amplitude obtained from the experiments and from 
the MSV calculation, all at 23.5 GeV.  }

\bigskip
 
{\small
  Fig. 15~-~ Real part  ${\cal K}(b)$ of the experimental profile 
functions for all ISR energies, 546 GeV and 1800 GeV.}

\bigskip

{\small
 Fig. 16~-~ Values of the experimental ratio 
$[b^{(3)}_{\rm max}]^2/\langle b^2\rangle$
 (black squares) and its mean value (line).} 

\bigskip
 
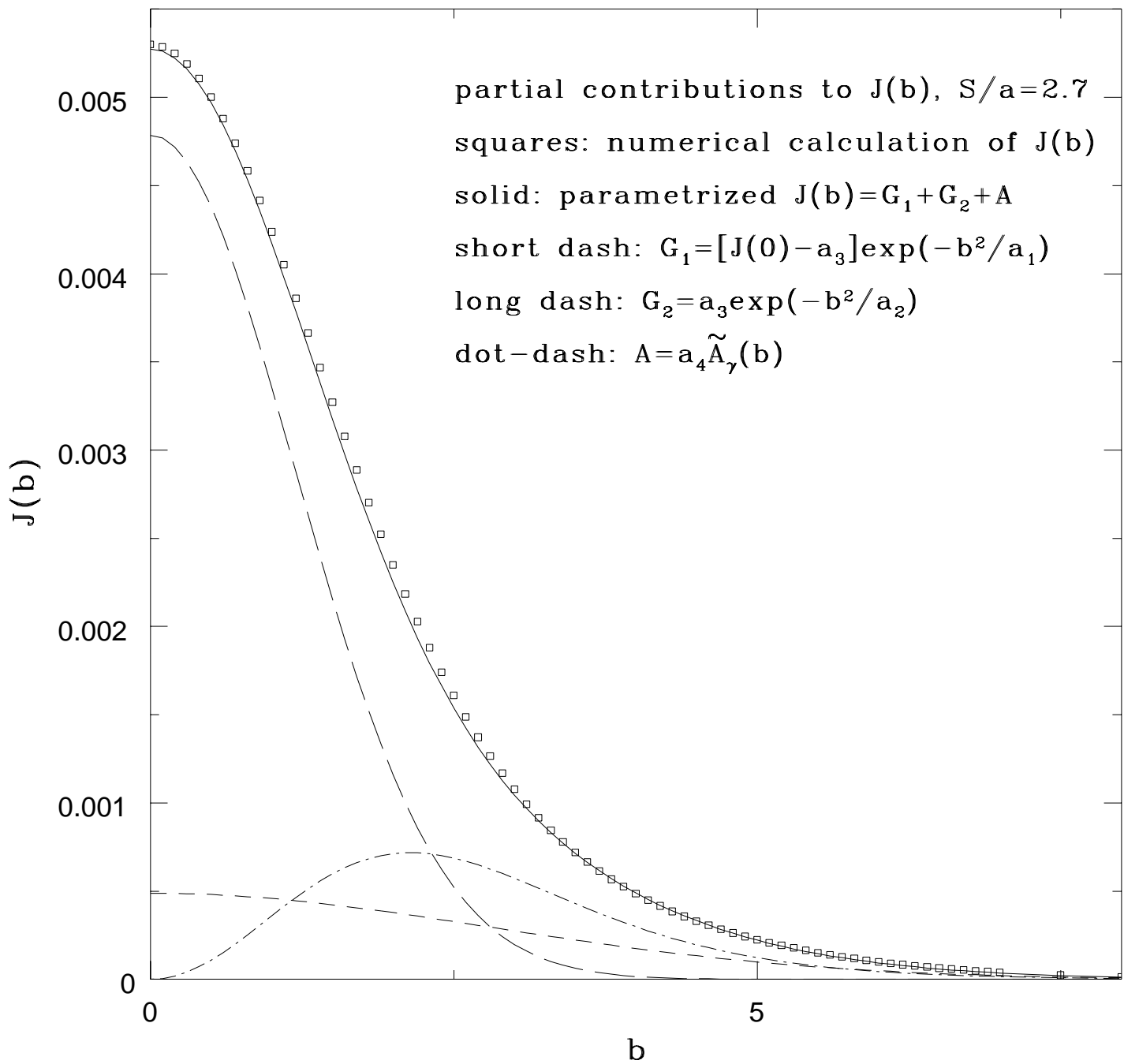
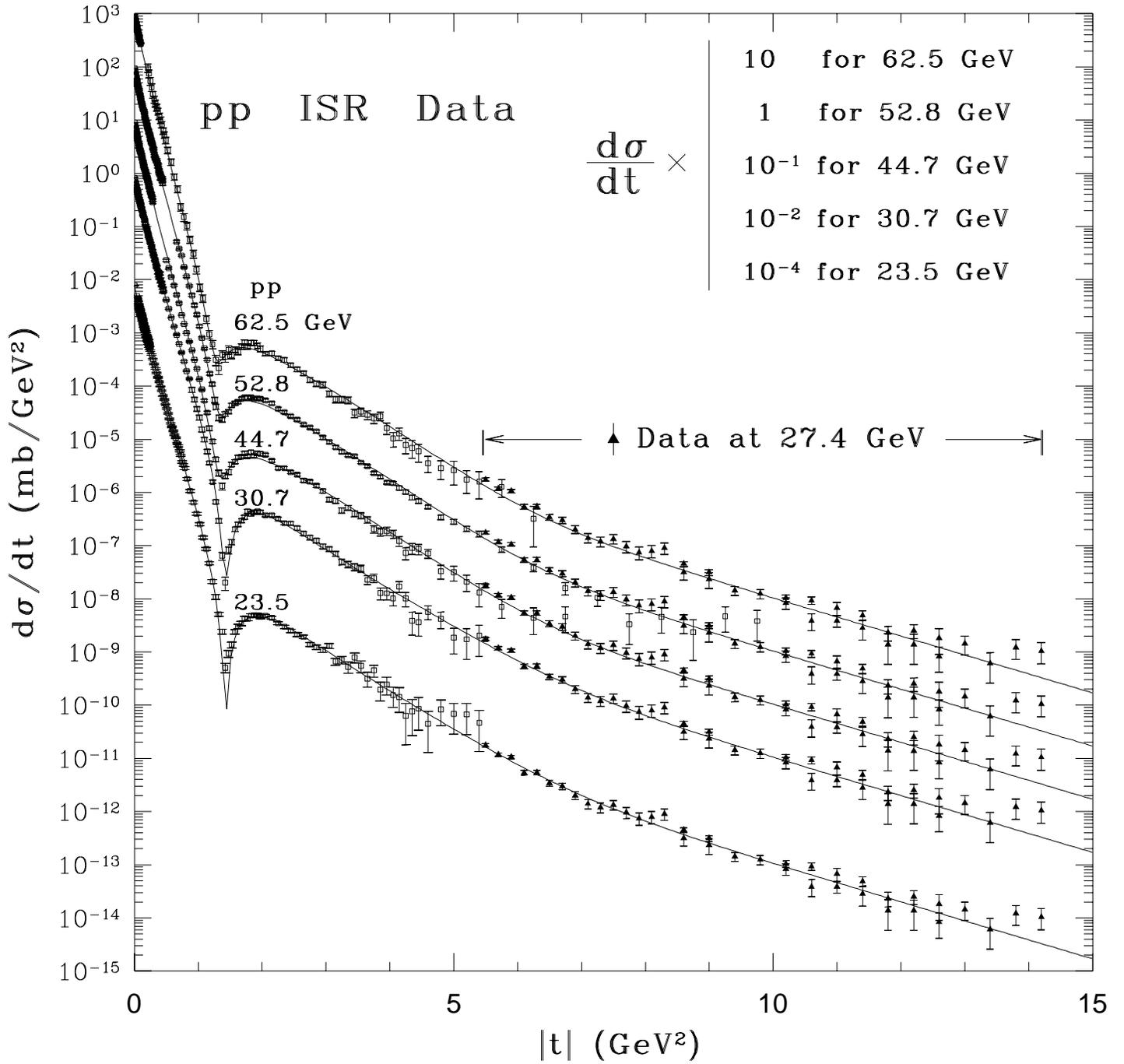
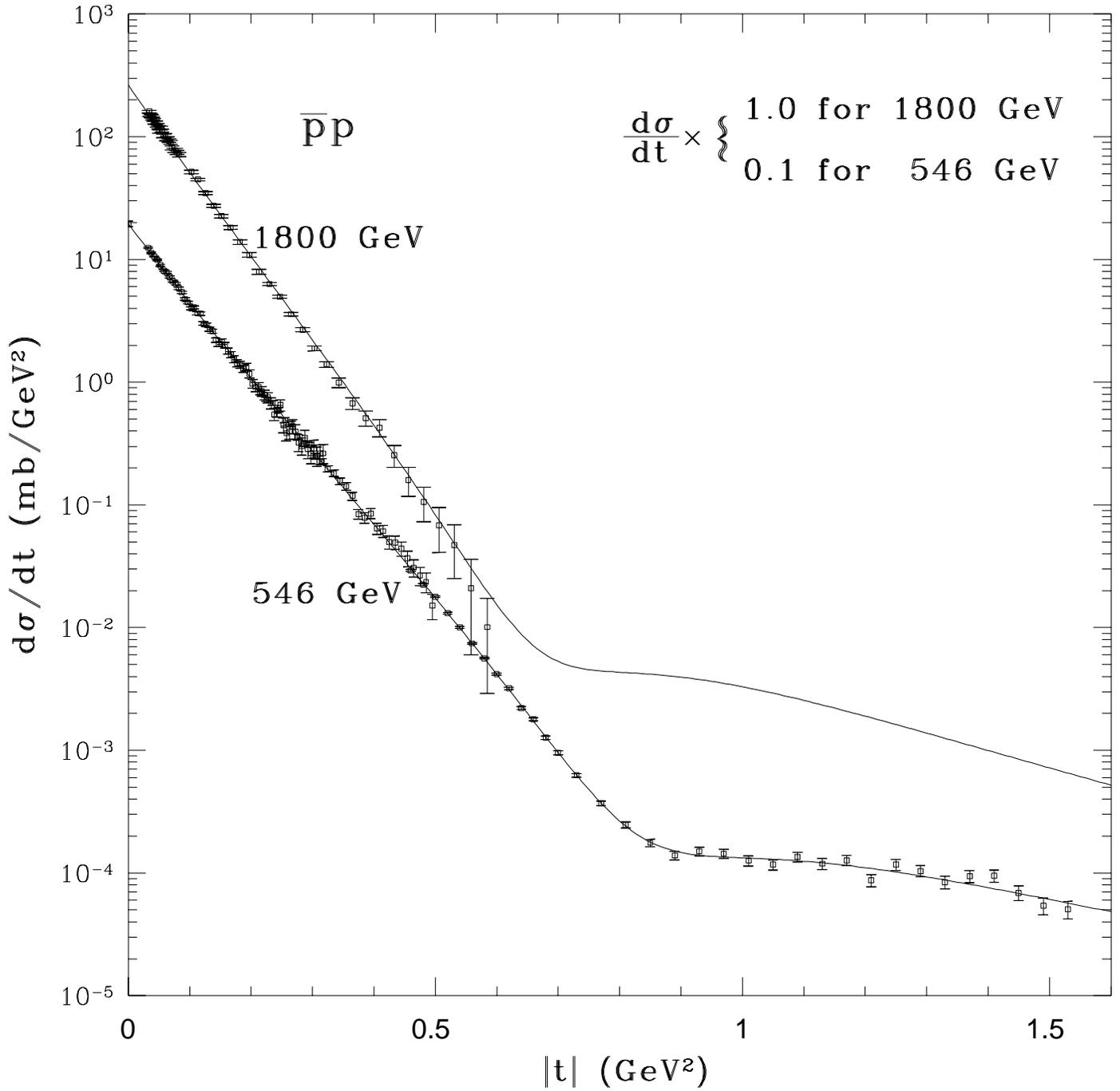
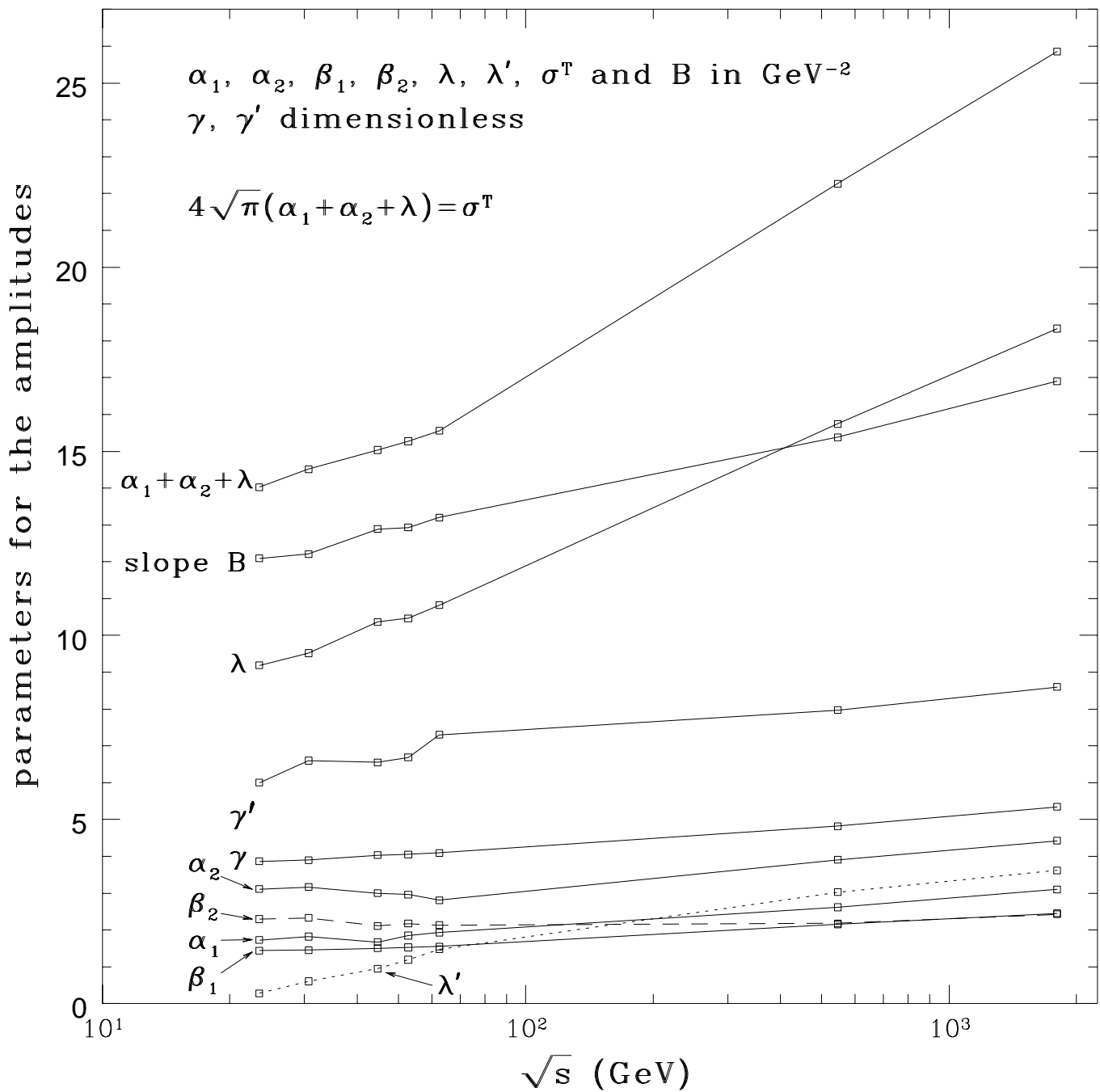
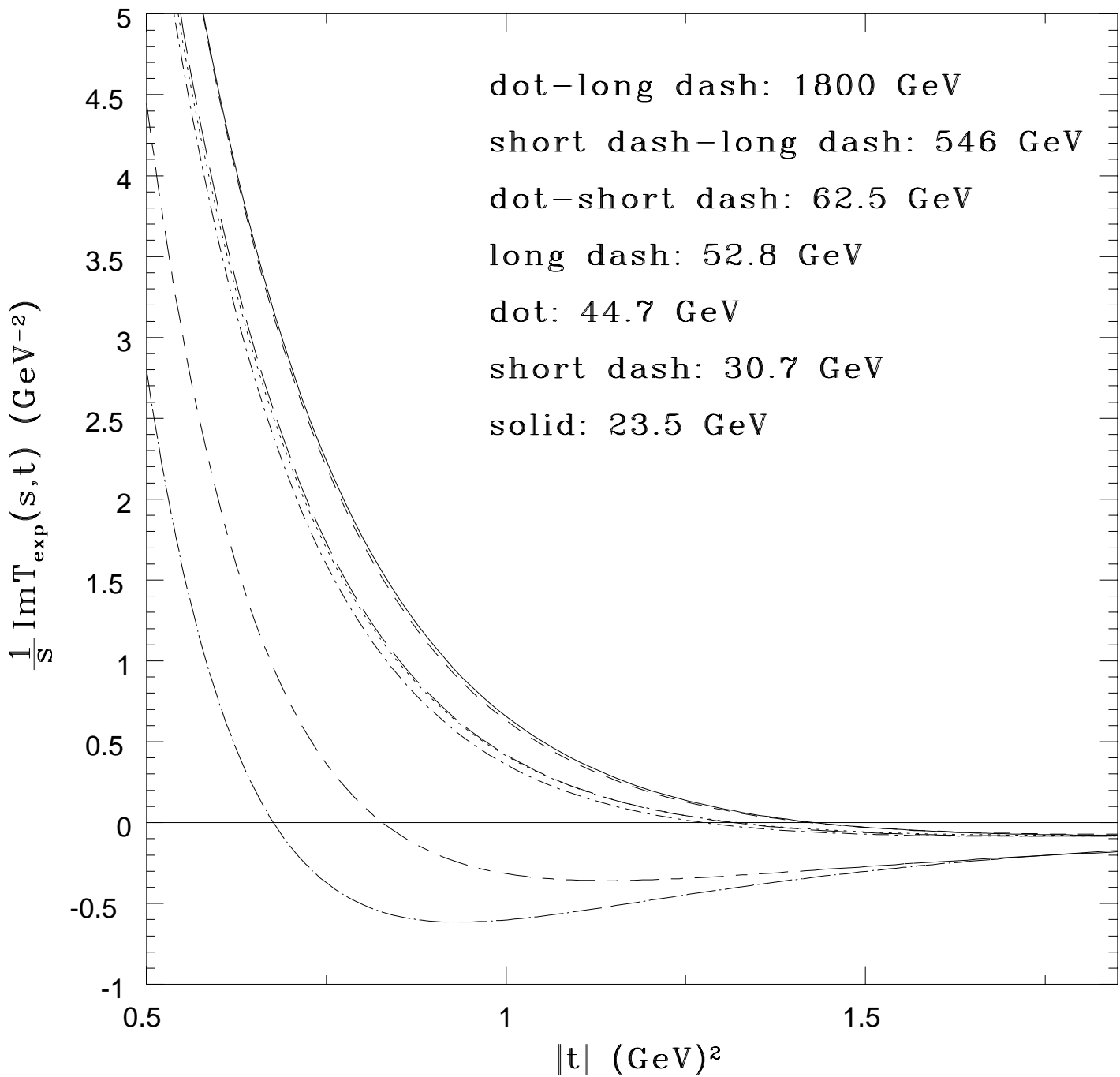
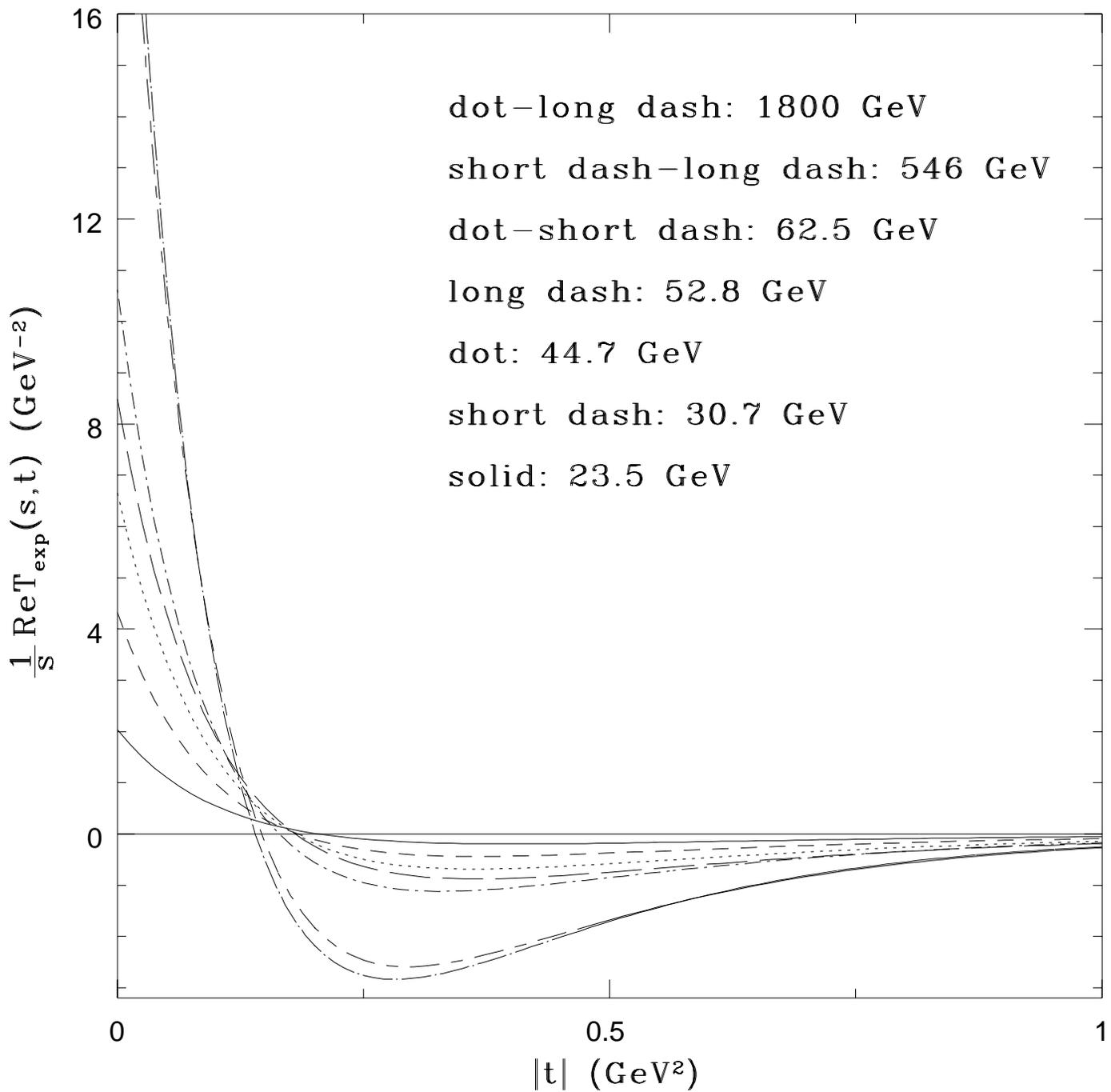
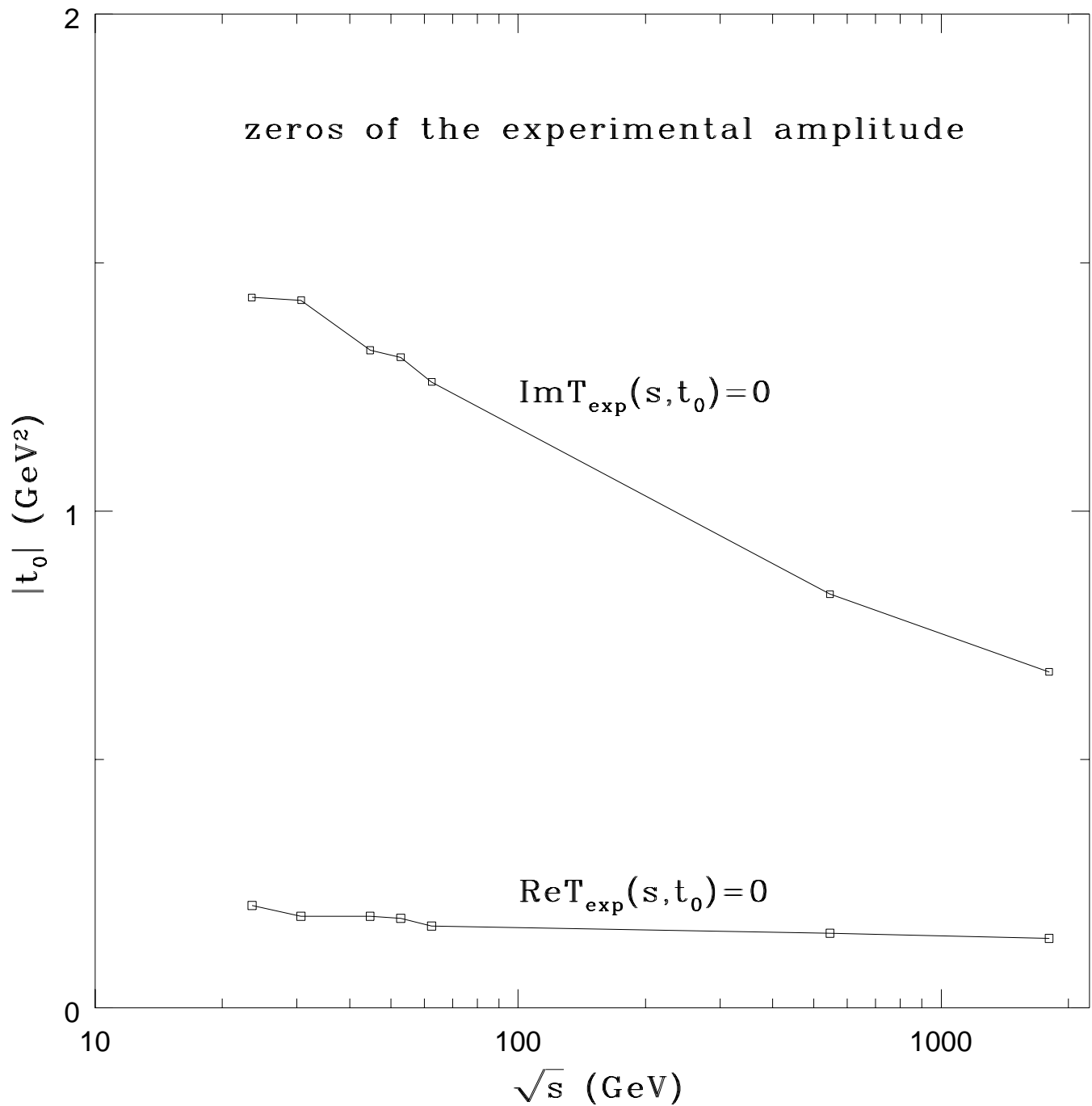
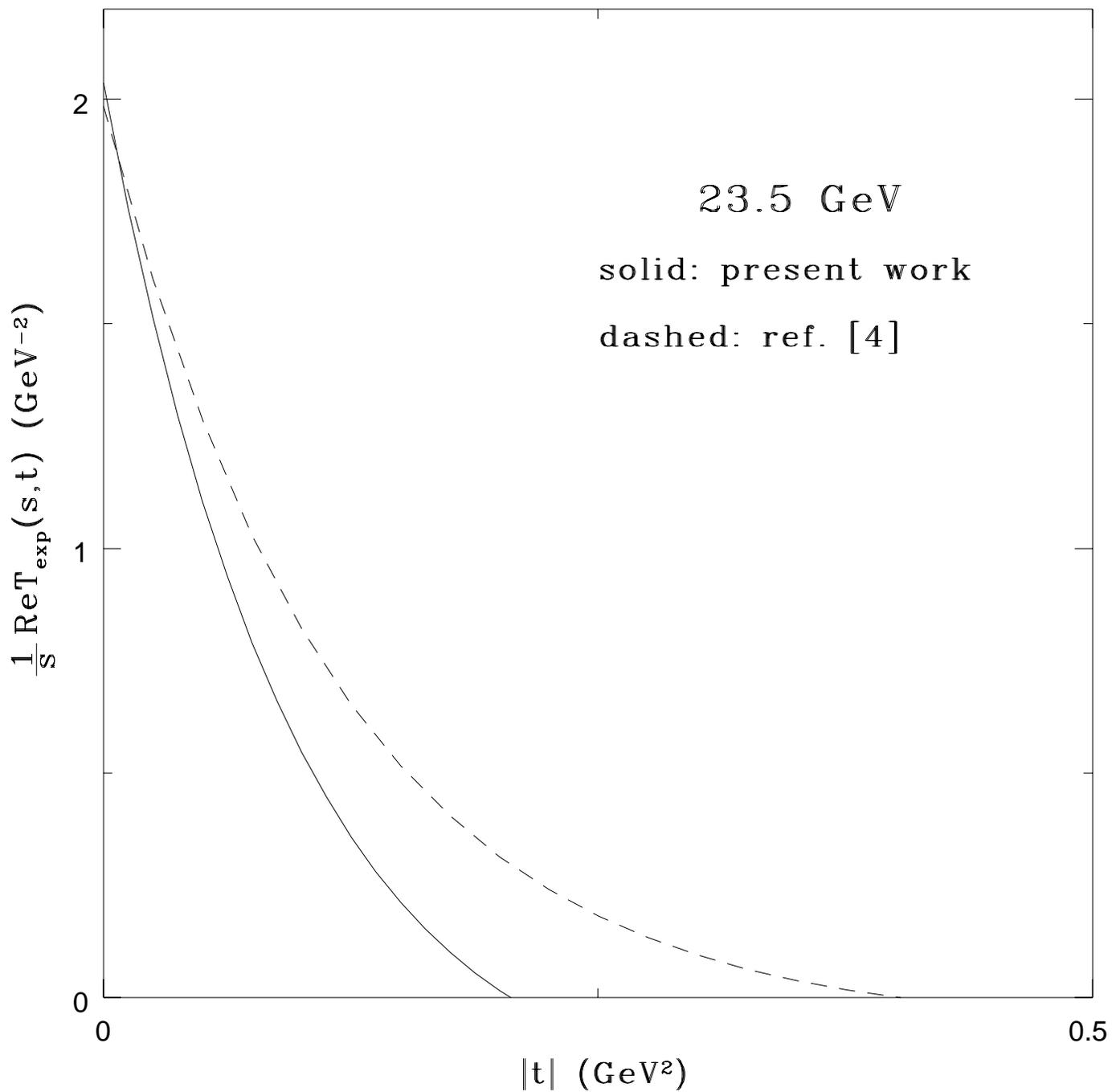
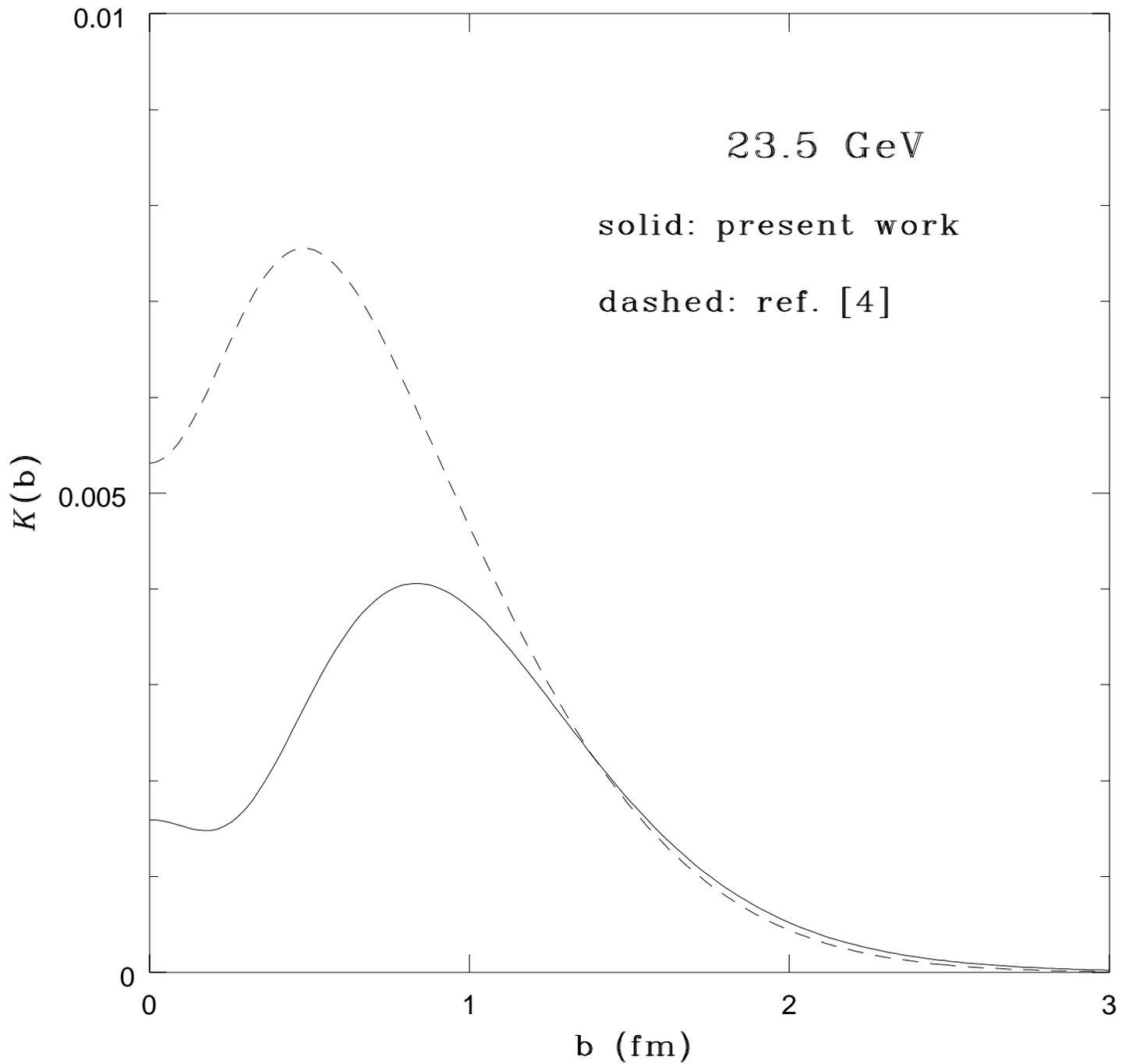
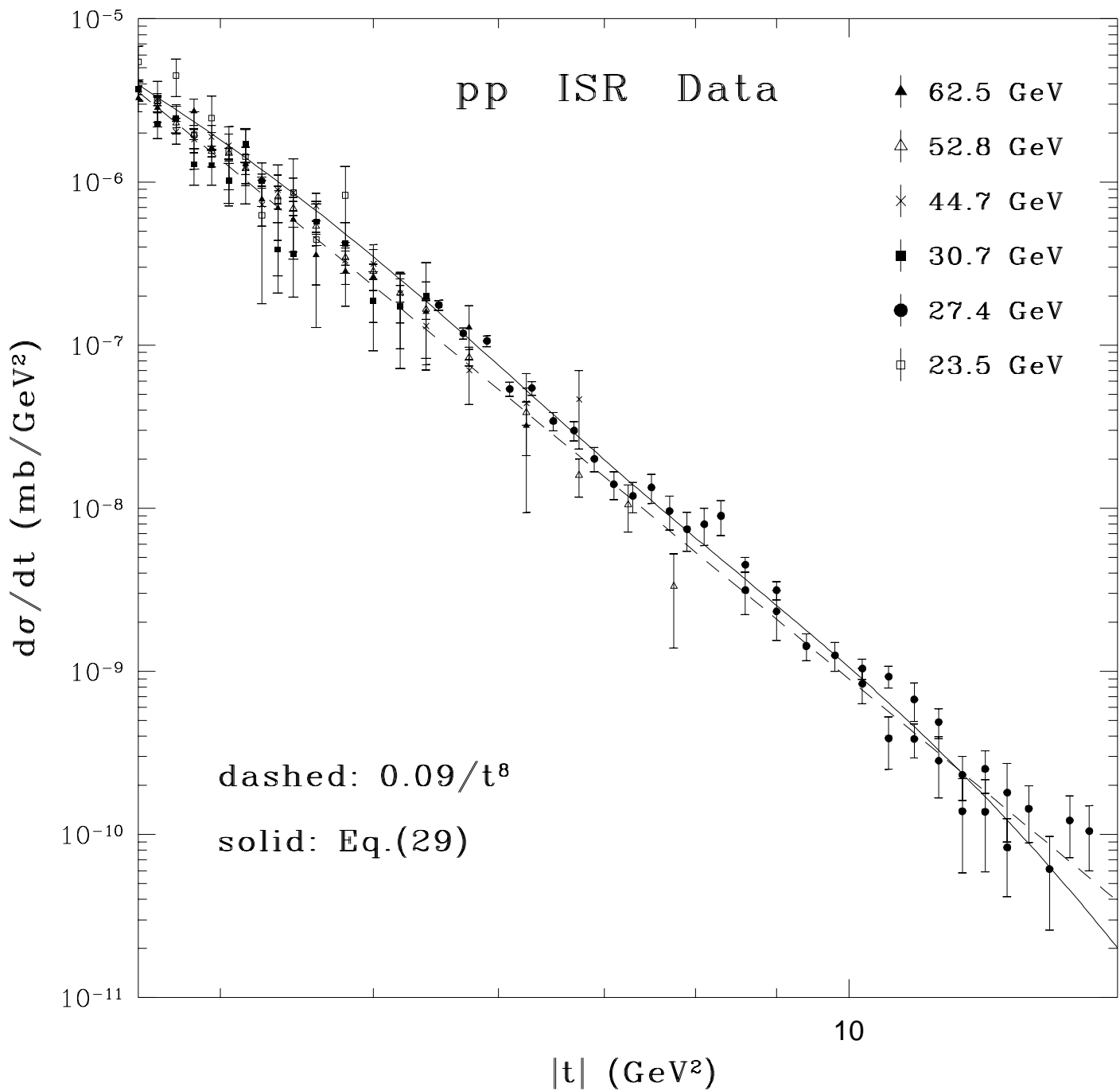
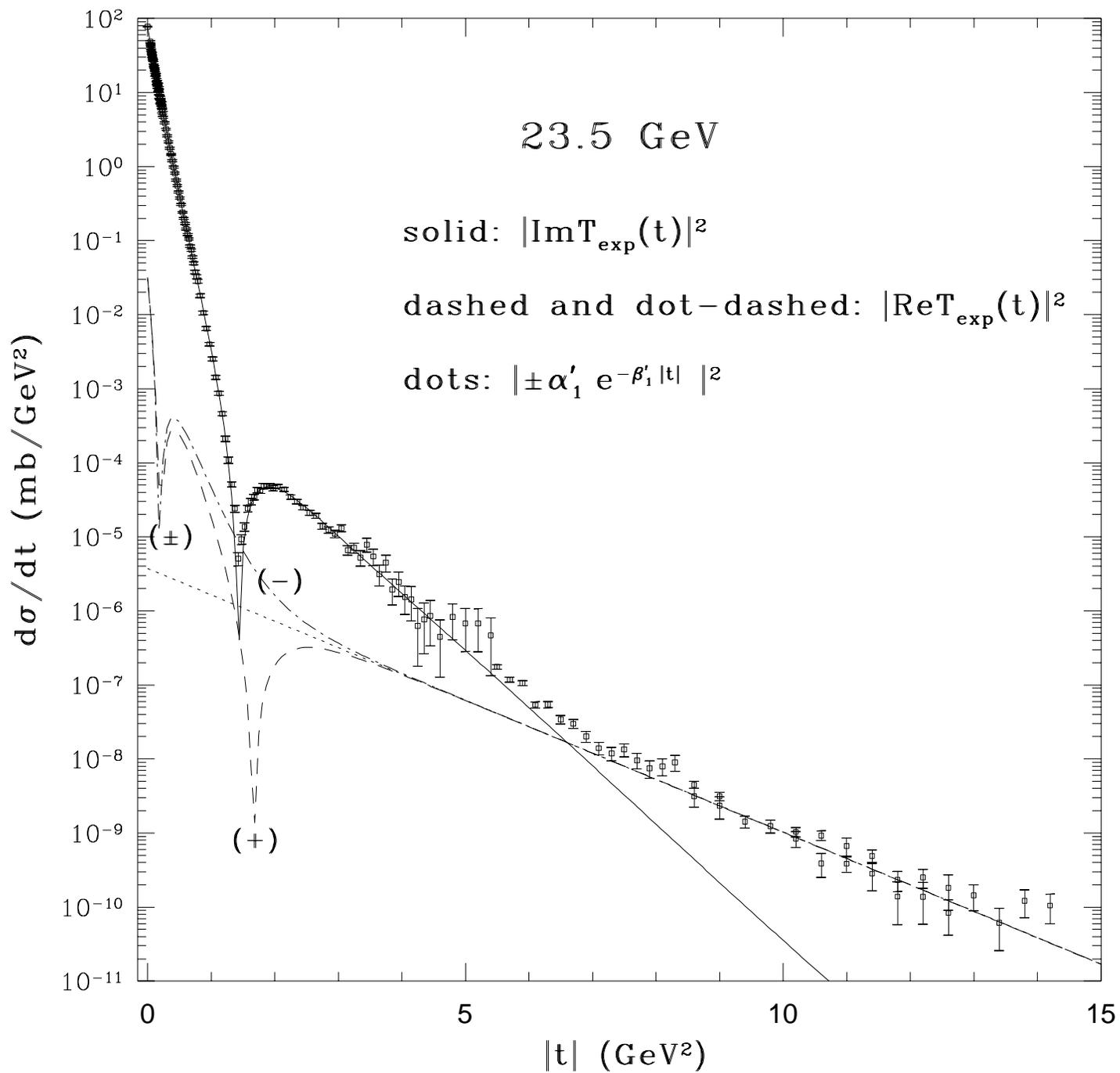
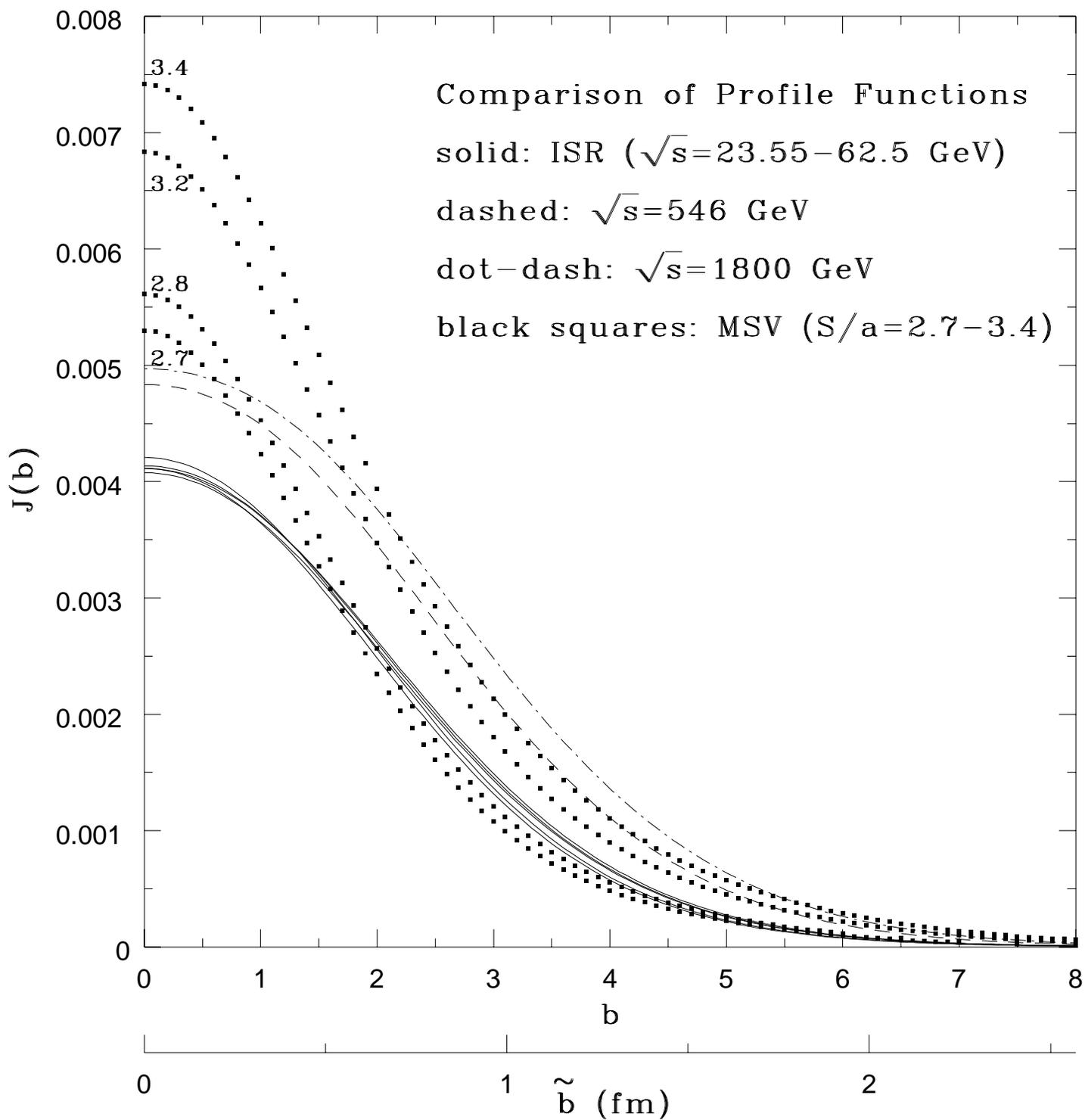
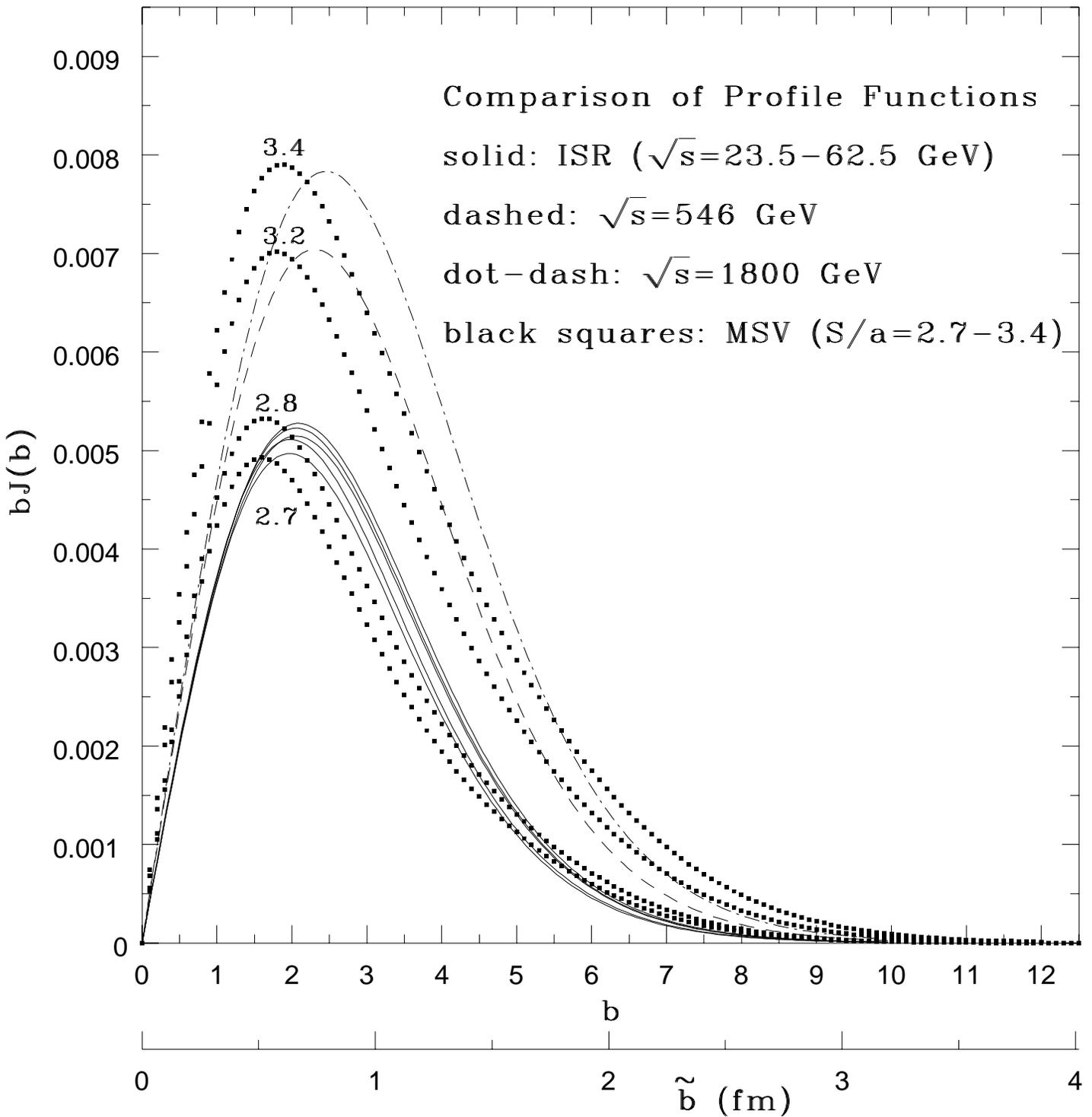
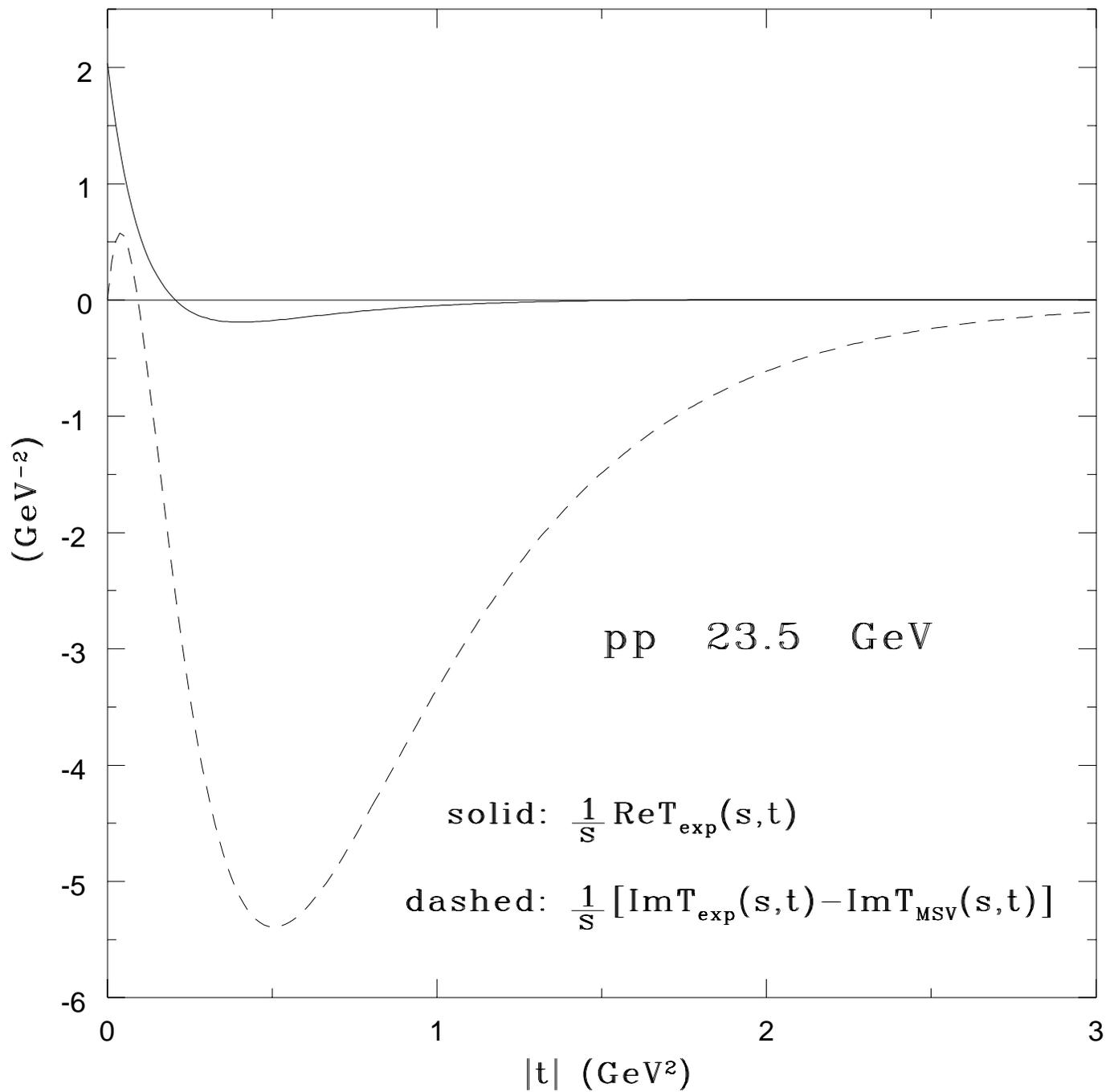
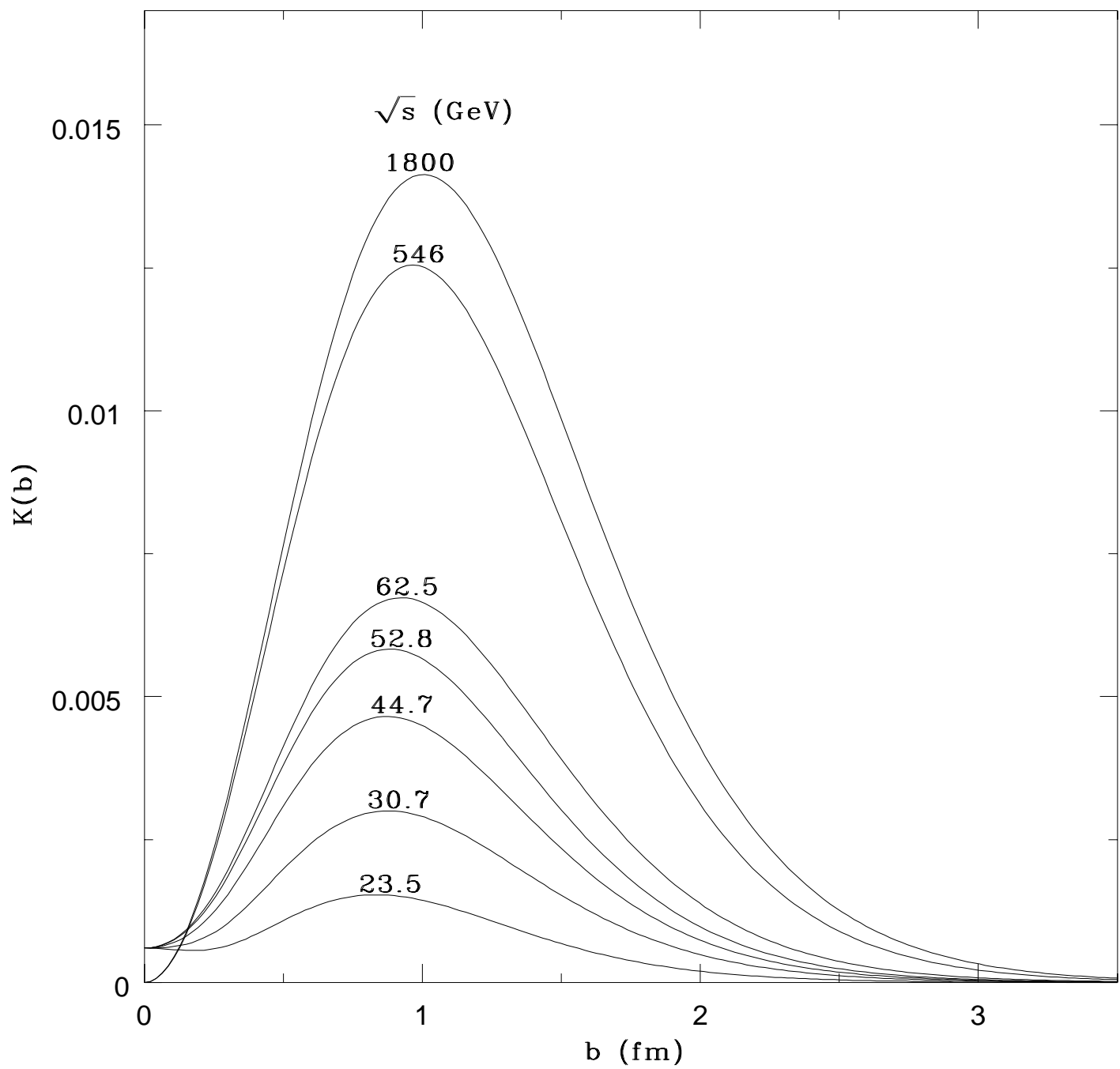
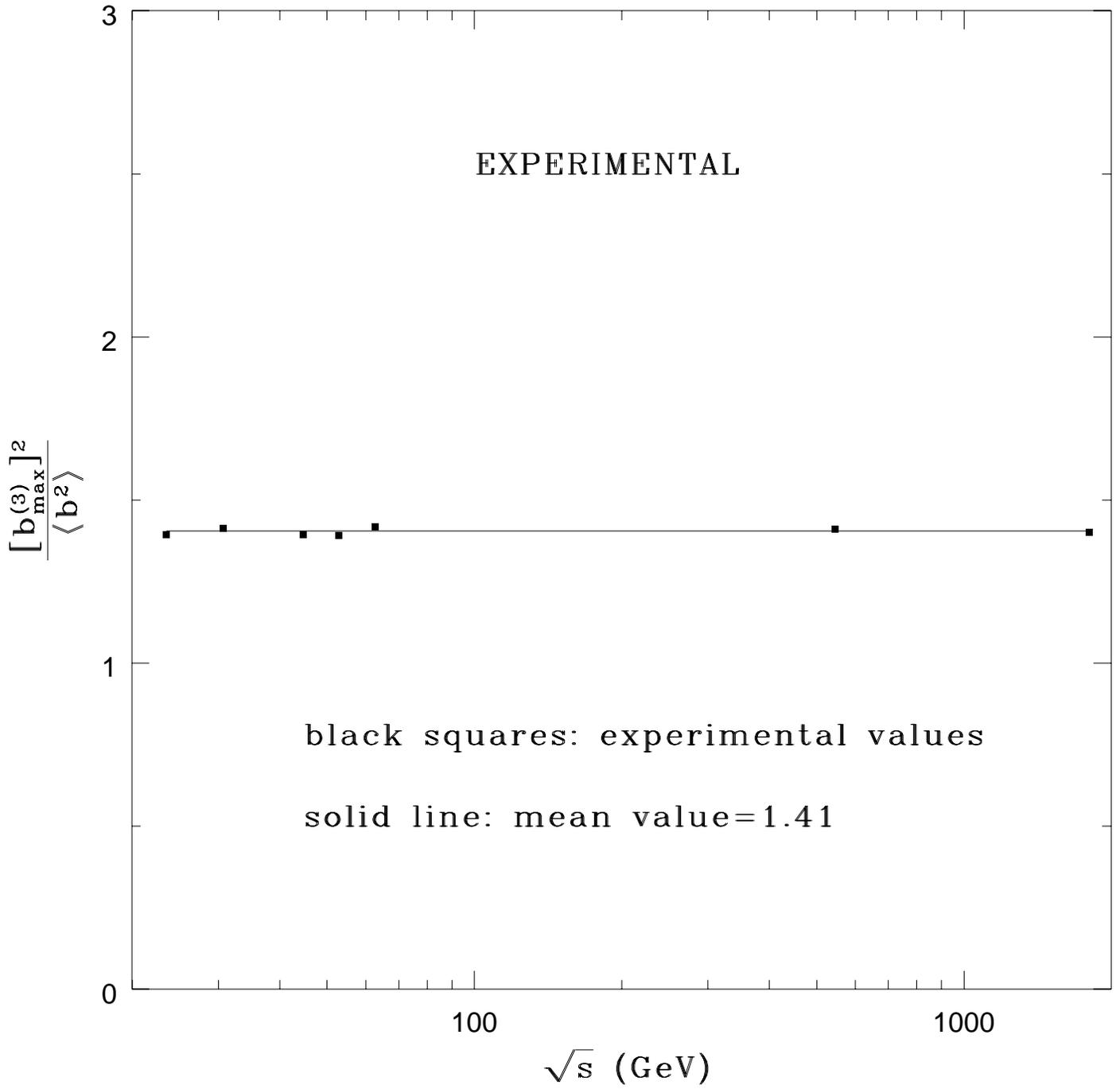
\end{document}